\documentclass[twocolumn,aps,pra,superscriptaddress]{revtex4-1}
\usepackage{graphicx}
\usepackage{amsmath}
\usepackage[normalem]{ulem} 
\usepackage{amssymb}
\usepackage{braket}
\usepackage{bm}
\usepackage[colorlinks=true,linkcolor=blue,citecolor=blue,urlcolor=blue]{hyperref}
\begin{document}


\title{Scattering and Bound States of Two Heteronuclear Ultracold Atoms in a Quasi-Two-Dimensional   Confinement}
\author{Binhao Wang}
\affiliation{School of Physics, Renmin University of China, Beijing, 100872, P. R. China}

%
%

 \author{Fan Yang}%
\affiliation{Hefei National Laboratory, Hefei 230088, China}

 \author{Peng Zhang}%
 \email{pengzhang@ruc.edu.cn}
\affiliation{School of Physics, Renmin University of China, Beijing, 100872, P. R. China}

\date{\today}

\begin{abstract}

We solve the two-body problem of ultracold heteronuclear atoms in a quasi-two-dimensional (quasi-2D) geometry. 
The quasi-2D confinement is realized by a harmonic trap along the longitudinal ($z$-) direction, with different trap frequencies for the two atoms, as in many current experiments on ultracold heteronuclear gases. 
As a consequence, the longitudinal center-of-mass (CoM) motion is coupled to the relative motion, which significantly complicates the two-body problem. 
We solve this problem exactly and derive the 2D scattering length $a_{\rm 2D}$, the 2D effective range parameter $R_{\rm 2D}$, and the bound-state energies, as functions of the $s$-wave scattering length and effective range of the two atoms in free three-dimensional (3D) space. We show that multiple 2D scattering resonances can be induced by the coupling between the longitudinal CoM and relative motion. 
Around these resonances, $a_{\rm 2D}$ varies rapidly with the 3D scattering parameters, while $R_{\rm 2D}$ is strongly enhanced. 
Since the effective pairwise interaction in quasi-2D ultracold gases is determined by i.e., the two-body scattering amplitudes and bound-state energies, our results can be used for manipulating the effective 2D interatomic interaction in quasi-2D ultracold heteronuclear gases by tuning the confinement frequencies and the 3D scattering parameters. 

\end{abstract}

\maketitle


\section{Introduction}
\label{sec:level1}

Quasi-two-dimensional (quasi-2D) ultracold gases provide ideal platforms for quantum simulation of two-dimensional quantum many-body physics \cite{2D,2DBKT1,2DBKT2,BoseBKT,Q2D1,Q2D2}.
 In recent years, substantial experimental and theoretical progress has been made in the study of these systems \cite{2DBose1,2DBose2,2DBose3,2DBose4,2DFeimi1,2DFeimi2,2DFeimi3,2DFeimi4,Petrov,Duan2006,Hu}, including, for example, the quantum simulation of superfluidity \cite{2DBose3} and Berezinskii--Kosterlitz--Thouless (BKT) physics in bosonic gases \cite{2DBKT1,2DBKT2,BoseBKT}, as well as two-dimensional pairing and the Bose--Einstein-condensation--Bardeen--Cooper--Schrieffer (BEC--BCS) crossover in two-component Fermi gases \cite{2DFeimi1,2DFeimi2,2DFeimi3,2DFeimi4}.
In parallel, heteronuclear ultracold atomic gases have attracted considerable attention in ultracold-atom physics, offering unique platforms for studying few- and many-body physics in mass-imbalanced two-species quantum systems \cite{heteronuclear1,heteronuclear2,heteronuclear3,heteronuclear4,heterFemi1,heterFemi2}. 
Important examples include Efimov physics and unconventional pairing in Fermi gases with mismatched Fermi surfaces \cite{heteronuclear2,heteronuclear3,heterFemi1,heterFemi2}. Considerable experimental and theoretical advances have been achieved in these directions over the past years.

At the intersection of these two research directions, quasi-2D heteronuclear ultracold atomic gases constitute a natural and promising platform for ultracold-atom physics \cite{heteronuclear2D}. 
Importantly, the key experimental conditions required for realizing such systems have already been established. 
It is therefore reasonable to expect that such systems will become experimentally accessible in the near future.

In quasi-2D heteronuclear ultracold atomic gases, the effective 2D pairwise interaction is determined by the solution of the corresponding two-body problem, namely, the two-body scattering amplitudes and bound-state energies \cite{solve2D1,solve2D2,2024}. 
Therefore, solving this two-body problem constitutes a necessary first step toward understanding quasi-2D heteronuclear ultracold atomic gases.

In this work, we focus on this two-body problem. 
Specifically, we consider two ultracold heteronuclear atoms confined in a quasi-two-dimensional (quasi-2D) geometry, as illustrated in Fig.~\ref{scheme}, where the longitudinal direction, i.e., the \(z\) direction, is subject to harmonic confinement. 
As in many current experiments with heteronuclear atoms, the two atoms experience different longitudinal confinement frequencies. 
This difference couples the center-of-mass (CoM) and relative motions along the longitudinal direction, thereby significantly complicating the two-body problem. 

We exactly solve this problem, and derive the 2D scattering length \(a_{\mathrm{2D}}\), the 2D effective range parameter \(R_{\mathrm{2D}}\), and the bound-state energy \(E_b\) for various values of the
longitudinal-confinement frequencies,
 three-dimensional (3D) scattering length \(a_{\mathrm{3D}}\) and effective range \(R_{\mathrm{3D}}\). 
We show that the coupling between the longitudinal CoM and relative motion can give rise to multiple 2D scattering resonances and bound states in this system. Around these resonances,  \(a_{\mathrm{2D}}\) and  \(R_{\mathrm{2D}}\) change rapidly. Our results therefore provide a possible means of controlling the effective 2D pairwise interaction by tuning the longitudinal confinement frequencies and  \(a_{\mathrm{3D}}\) and \(R_{\mathrm{3D}}\), for example via a magnetic Feshbach resonance.

The remainder of this paper is organized as follows. In Sec.~\ref{sec:level2}
we introduce our system and the to-be-calculated quantities in detail.
The details of our calculation approach are shown in the appendixes.
In Sec.~\ref{res}, we present our results and discuss the mechanism underlying the resonances found in our calculations. A summary of this work is given in Sec.~\ref{summary}.


\section{Two Heteronuclear Atoms In a Quasi-2D Confinement}
\label{sec:level2}

\begin{figure}[t]
    \centering
    \includegraphics[width=0.75\linewidth]{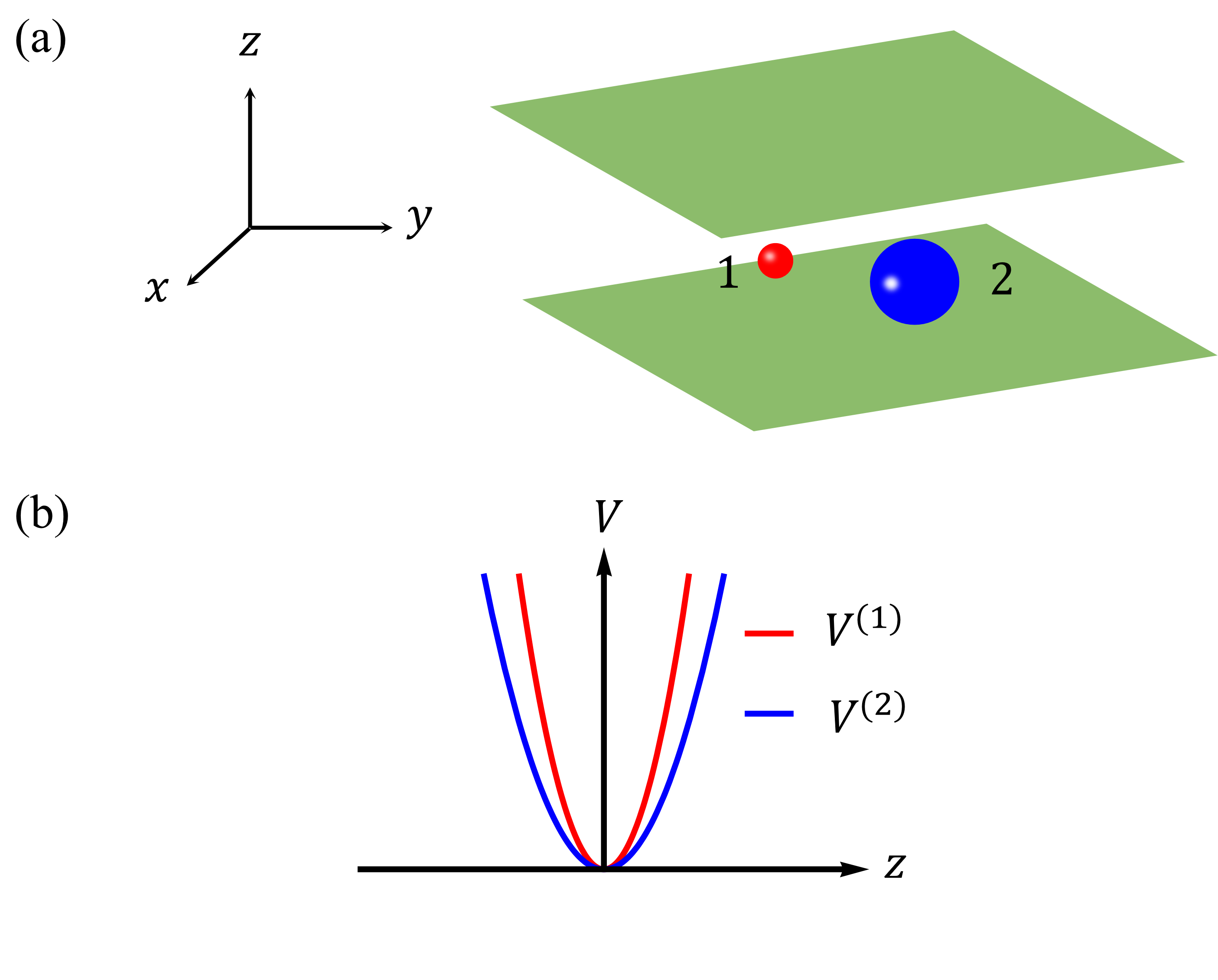}
    \caption{\textbf{(a)}: Schematic diagram of two heteronuclear  ultracold   atoms in a quasi-2D  confinement, which is realized via a strong  confinement  along the  longitudinal ($z$-) direction. \textbf{(b)}: Schematic diagram of the harmonic   longitudinal confinement potentials $V^{(1)}$ and $V^{(2)}$, experienced by atoms 1 and 2, respectively. }
    \label{scheme}
  \end{figure}

We consider two ultracold heteronuclear atoms, labeled as atoms 1 and 2, confined by harmonic potentials along the \(z\) direction, as illustrated in Fig.~\ref{scheme}. 
As mentioned above, the two atoms experience longitudinal confinements with different trap frequencies. 
In the \(x\)-\(y\) plane, no confinement is applied. 
For this system, the CoM motion in the \(x\)-\(y\) plane can be separated out. 
As a result, the relevant degrees of freedom are reduced to the relative motion in all three spatial directions and the CoM motion only along the \(z\) direction. 
The corresponding Hamiltonian is given by, with \(\hbar=1\),
\begin{eqnarray}
    \hat{H} &=& -\frac{1}{2\mu}\nabla_{\bm{r}}^2  
    -\frac{1}{2M}\frac{\partial^2}{\partial Z^2}
    +\sum_{j=1,2}V^{(j)}(z_j)+U(\bm{r}).
    \label{h}
\end{eqnarray}
Here 
\(M=m_1+m_2\) and \(\mu=m_1m_2/(m_1+m_2)\) are the total and reduced masses of the two atoms, respectively, where \(m_j\) \((j=1,2)\) is the mass of atom \(j\).
Moreover, \({\bm r}=(x,y,z)\) denotes the relative coordinate, and 
\(Z=(m_1z_1+m_2z_2)/(m_1+m_2)\) is the longitudinal CoM coordinate, with \(z_j\) being the longitudinal coordinate of atom \(j\). 
The longitudinal confinement potential for atom \(j\) is
\begin{equation}
V^{(j)}(z_j) = \frac{1}{2}m_j\omega_j^2z_j^2,
\label{eq.curve}
\end{equation}
where \(\omega_j\) is the corresponding angular trap frequency [Fig.~\ref{scheme}(b)].
Additionally, in Eq.~(\ref{h}), \(U(\bm{r})\) is an energy-dependent zero-range pseudopotential~\cite{LHY1,LHY} describing the interatomic interaction. 
It is characterized by the \(s\)-wave scattering length \(a_{\rm 3D}\) and effective range \(R_{\rm 3D}\) between the two atoms in free three-dimensional (3D) space, and can be expressed as
\begin{eqnarray}
 U(\bm{r}) &\equiv&\frac{2\pi \hat{A}_{\rm 3D}
 }{\mu}\delta(\bm{r})  \frac{\partial}{\partial r} (r\cdot).
 \label{uhy}
\end{eqnarray}
Here \(\hat{A}_{\rm 3D}\) is an operator acting on the Hilbert space of the longitudinal CoM motion and encodes the dependence on the 3D scattering parameters \(a_{\rm 3D}\) and \(R_{\rm 3D}\). When \(R_{\rm 3D}=0\), we have \(\hat{A}_{\rm 3D}=a_{\rm 3D}\). 
The explicit expression of \(\hat{A}_{\rm 3D}\) is given in Appendix~\ref{a3d}.
To facilitate our analysis, we  further define $\phi^{(j)}_n(z_j)$ $(j=1,2; n=0,1,2,...)$ as the normalized eigenfunctions of the longitudinal Schrödinger equation for atom $j$, i.e.,
\begin{align}
 \left[-\frac{1}{2m_j}\frac{\partial^2}{\partial z_j^2} + V^{(j)}(z_j)\right]\phi^{(j)}_n(z_j) &= \left(n+\frac{1}{2}\right)\omega_j\phi^{(j)}_n(z_j), \nonumber \\
 &(j=1,2; n=0,1,2,...).
\end{align}

Now we consider the scattering between atoms 1 and 2, both initially prepared in the  longitudinal ground state. The incident state of this scattering process is
\begin{eqnarray}
\Psi_{\bm k}^{\rm(in)}({\bm \rho},z_1,z_2)=\frac{1}{2\pi}e^{i{\bm k}\cdot{\bm \rho}}\phi_0^{(1)}(z_1)\phi_0^{(2)}(z_2),\label{incident}
\end{eqnarray}
where ${\bm \rho}=(x,y)$, and ${\bm k}$ is the incident momentum in the $x$-$y$ plane.
During the scattering process, the inter-atomic interaction $U(\bm{r})$ can couple different longitudinal states $\phi^{(1)}_n(z_1)\phi^{(2)}_m(z_2) $ $(m,n=0,1,2,...)$ with each other. As a result, the corresponding scattering wave function $\Psi_{\bm k}^{(+)}(\bm{\rho}, z_1, z_2)$ for this incident state incorporates both the ground and excited longitudinal states, and thus can be expressed as:
\begin{eqnarray}
\Psi_{\bm k}^{(+)}({\bm \rho}, z_1, z_2)= \sum_{m, n=0}^{\infty}\phi_m^{(1)}(z_1)\phi_n^{(2)}(z_2)\psi_{\bm k}^{(m,n)}({\bm \rho}).\label{psiexp0}
\end{eqnarray}
Here, we restrict our analysis to the low-energy regime where the relative kinetic energy satisfies $k^2/(2\mu) \ll \omega_{1,2}$. Under this condition, all contributions from  longitudinal excited states exhibit exponential decay at large interatomic separations. Consequently, the components $\psi_{\bm k}^{(m,n)}(\bm{\rho})$ satisfy the long-range boundary conditions:
\begin{eqnarray}
\lim_{\rho\rightarrow \infty}\psi_{\bm k}^{(m,n)}({\bm \rho})&=&0,\ \ \ \ \ \ \ \ \ \ \ \ (m,n)\neq (0,0),
\label{con-1} \\[5pt]
\lim_{\rho \rightarrow \infty}\psi_{\bm k}^{(0,0)}({\bm \rho})&=&\frac{1}{2\pi}\left[e^{i{\bm k}\cdot{\bm \rho}}\!+\!f_{\rm 2D}(k)\frac{e^{ik\rho}}{\sqrt{\rho}}\right],\label{con0}
\end{eqnarray}
with $\rho=|{\bm \rho}|$ and $k=|{\bm k}|$. As a result,
 in the limit $\rho\rightarrow\infty$
 the total scattering wave function $\Psi_{\bm k}^{(+)}({\bm \rho}, z_1, z_2)$ satisfies
 \begin{eqnarray}
\lim_{\rho \rightarrow \infty}\!\!\Psi_{\bm k}^{(+)}\!\!\left({\bm \rho},\! z_1, \! z_2\right)\!=\!\frac{1}{2\pi}\!\!\left[e^{i{\bm k}\cdot{\bm \rho}}\!+\!f_{\rm 2D}(k)\frac{e^{ik\rho}}{\sqrt{\rho}}\right]\!\!\phi_0^{(1)}(z_1)\phi_0^{(2)}(z_2).\nonumber\\
\label{bc1}
\end{eqnarray}
Here $f_{\rm 2D}(k)$ is proportional to the effective 2D scattering amplitude, and can be expressed as
\begin{eqnarray}
f_{\rm 2D}(k)\approx \sqrt{\frac{\pi}{2k}}e^{i\pi/4}\frac{-1}{i\frac{\pi}2-\gamma-\ln\left(k\frac{a_{\rm 2D}}{2}\right)
-\frac{1}{2}R_{\rm 2D}k^2}
,\label{f2D}
\end{eqnarray}
in the low-energy limit, where $\gamma\approx 0.577$ is the Euler's constant.
Here $a_{\rm 2D}$
and $R_{\rm 2D}$ are the 2D scattering length and effective range parameter, respectively. They satisfy:
\begin{eqnarray}
a_{\rm 2D}>0;\ \ \ R_{\rm 2D}<0.
\end{eqnarray}
Both $a_{\rm 2D}$ and $R_{\rm 2D}$
 are functions of the 3D scattering parameters $a_{\rm 3D}$ and $R_{\rm 3D}$, and  the  longitudinal confinement frequencies 
 $\omega_{1,2}$. We further define characteristic length $\ell_z$ as: 
\begin{eqnarray}
\ell_z= \sqrt{\frac{1}{m_1\omega_1}}.\label{l1mt}
\end{eqnarray}
A direct dimensional analysis yields that, the non-dimensionalized results  $a_{\rm 2D}/\ell_z$ and $R_{\rm 2D}/\ell_z^2$ are universal functions of the 
dimensionless parameters
$a_{\rm 3D}/\ell_z$, $R_{\rm 3D}/\ell_z$, $m_1/m_2$ and $\omega_1/\omega_2$, i.e.,
$a_{\rm 2D}/\ell_z$ and $R_{\rm 2D}/\ell_z^2$ can be formally expressed as
\begin{eqnarray}
\frac{a_{\rm 2D}}{\ell_z}&=&{\cal F}_a\left(\frac{a_{\rm 3D}}{\ell_z},\frac{R_{\rm 3D}}{\ell_z};\frac{m_1}{m_2},\frac{\omega_1}{\omega_2}\right);\\[8pt]
\frac{R_{\rm 2D}}{\ell_z^2}&=&{\cal F}_R\left(\frac{a_{\rm 3D}}{\ell_z},\frac{R_{\rm 3D}}{\ell_z};\frac{m_1}{m_2},\frac{\omega_1}{\omega_2}\right),
\end{eqnarray}
with ${\cal F}_a$ and ${\cal F}_R$ being universal functions.
Additionally, note that the effective 2D interaction between these two atoms {\color{blue}is} described by $a_{\rm 2D}$ and $R_{\rm 2D}$. Thus, one can tune this effective 2D interaction by changing the confinement frequencies $\omega_{1,2}$ and the 3D scattering parameters $a_{\rm 3D}$ and $R_{\rm 3D}$.

In addition to the scattering amplitude, 
another fundamental aspect of this system is the two-body bound state energy. In the present configuration, there is at least one two-body bound state. Similar to the 2D scattering parameters $a_{\rm 2D}$ and $R_{\rm 2D}$, 
 non-dimensionalized bound-state energy  $E_{b}\ell_z^2\mu$ is a universal function  of the 
dimensionless parameters
$a_{\rm 3D}/\ell_z$, $R_{\rm 3D}/\ell_z$, $m_1/m_2$ and $\omega_1/\omega_2$, i.e.,
we have
\begin{eqnarray}
E_{b}\ell_z^2\mu&=&{\cal F}_b\left(\frac{a_{\rm 3D}}{\ell_z},\frac{R_{\rm 3D}}{\ell_z};\frac{m_1}{m_2},\frac{\omega_1}{\omega_2}\right),
\end{eqnarray}
where ${\cal F}_b$ is a universal function. 
Therefore, the bound state energy can be controlled by  $\omega_{1,2}$, and $a_{\rm 3D}$ and $R_{\rm 3D}$.

We derive $(a_{\mathrm{2D}}, R_{\rm 2D}, E_{b})$, i.e., the universal functions ${\cal F}_a$, ${\cal F}_R$ and ${\cal F}_b$, via exact calculations. 
Our calculation approach  is directly generalized from our previous work~\cite{2024}, and  we present the details in Appendices~\ref{apps} and \ref{appbs}.
 In the next section, we  present the results and perform a detailed analysis.

\section{Results}
\label{res}

In this section we present the 2D scattering parameters, \(a_{\rm 2D}\) and \(R_{\rm 2D}\), and the bound-state energy \(E_b\) obtained from our calculation. 
As an illustrative example, we consider a system 
with atoms 1 and 2 being a \(^{6}\mathrm{Li}\) atom and a \(^{53}\mathrm{Cr}\) atom, respectively.  
Thus, the mass ratio is fixed as
\begin{eqnarray}
m_1:m_2=6:53.
\end{eqnarray}


\subsection{2D Scattering Parameters: $a_{\rm 2D}$ and $R_{\rm 2D}$.}
\label{2dsp}

\begin{figure}[t]
    \centering
    \includegraphics[width=1.1\linewidth]{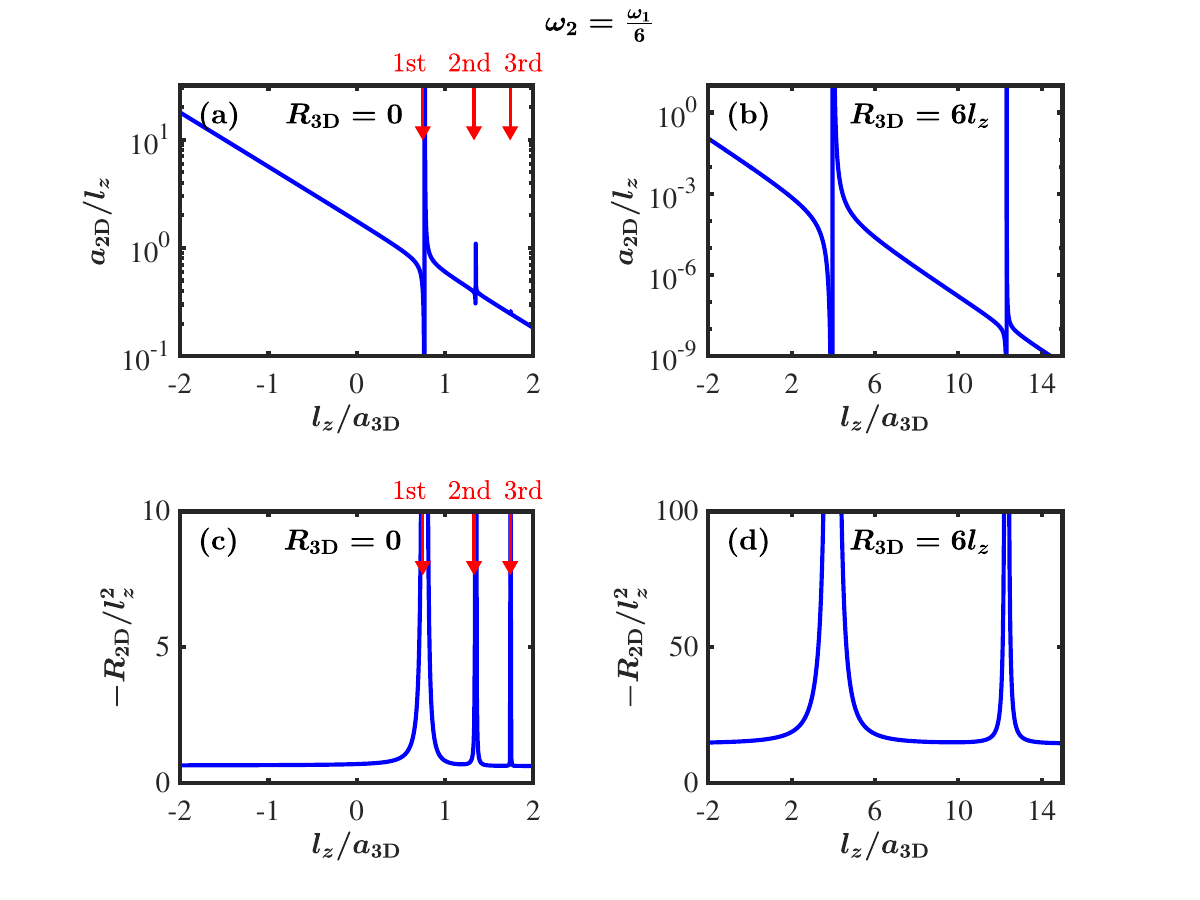}
    \includegraphics[width=1.1\linewidth]{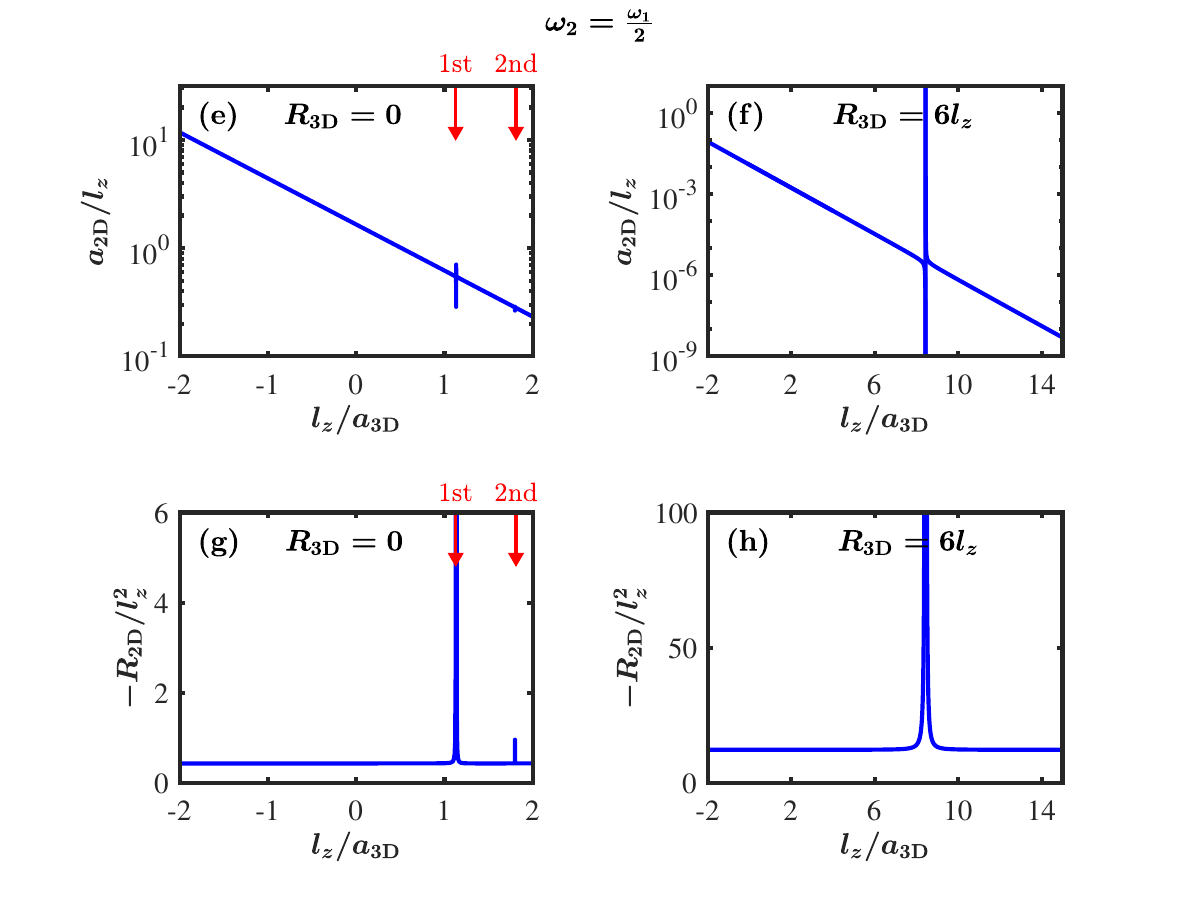}
        \caption{The 2D scattering length \(a_{\rm 2D}\) and effective range parameter
\(R_{\rm 2D}\) for a \(^{6}\mathrm{Li}\)-\(^{53}\mathrm{Cr}\) system
(\(m_1:m_2=6:53\)), in the case \(\omega_1>\omega_2\).
In panels (a--d) and (e--h) show the results for
\(\omega_1:\omega_2=6:1\) and \(\omega_1:\omega_2=2:1\), respectively.
For both frequency ratios, the results are presented as functions of
\(\ell_z/a_{\rm 3D}\), with \(R_{\rm 3D}=0\) in panels (a,c,e,g) and
\(R_{\rm 3D}=6\ell_z\) in panels (b,d,f,h). Here
\(\ell_z=\sqrt{1/(m_1\omega_1)}\), as defined in Eq.~(\ref{l1mt}).
In panels (a,c,e,g), the red arrows indicate the resonance positions
estimated from Eq.~(\ref{con}). 
For details, see Sec.~\ref{2dsp}.
        }

    \label{Fig1}
  \end{figure}
  
  \begin{figure}[t]
    \centering
    \includegraphics[width=1.1\linewidth]{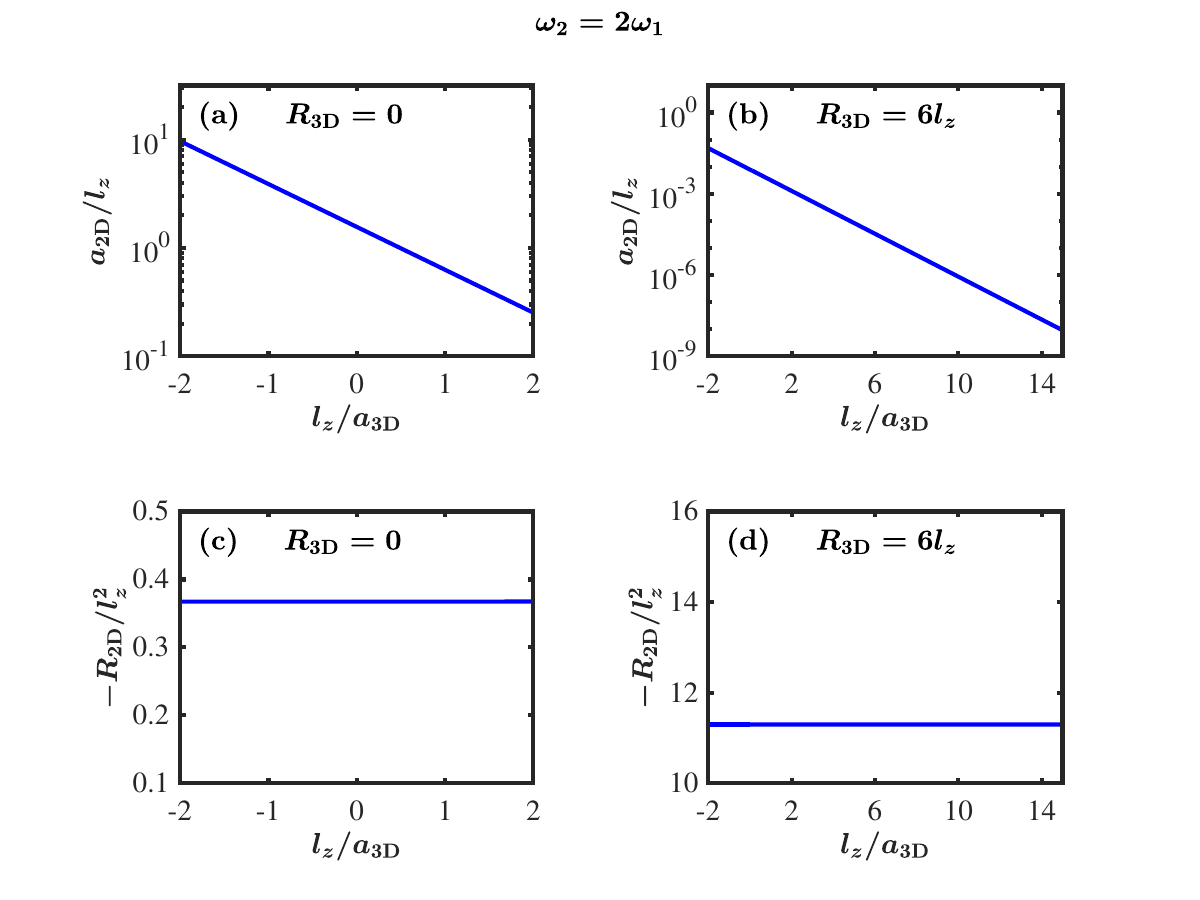}
    \includegraphics[width=1.1\linewidth]{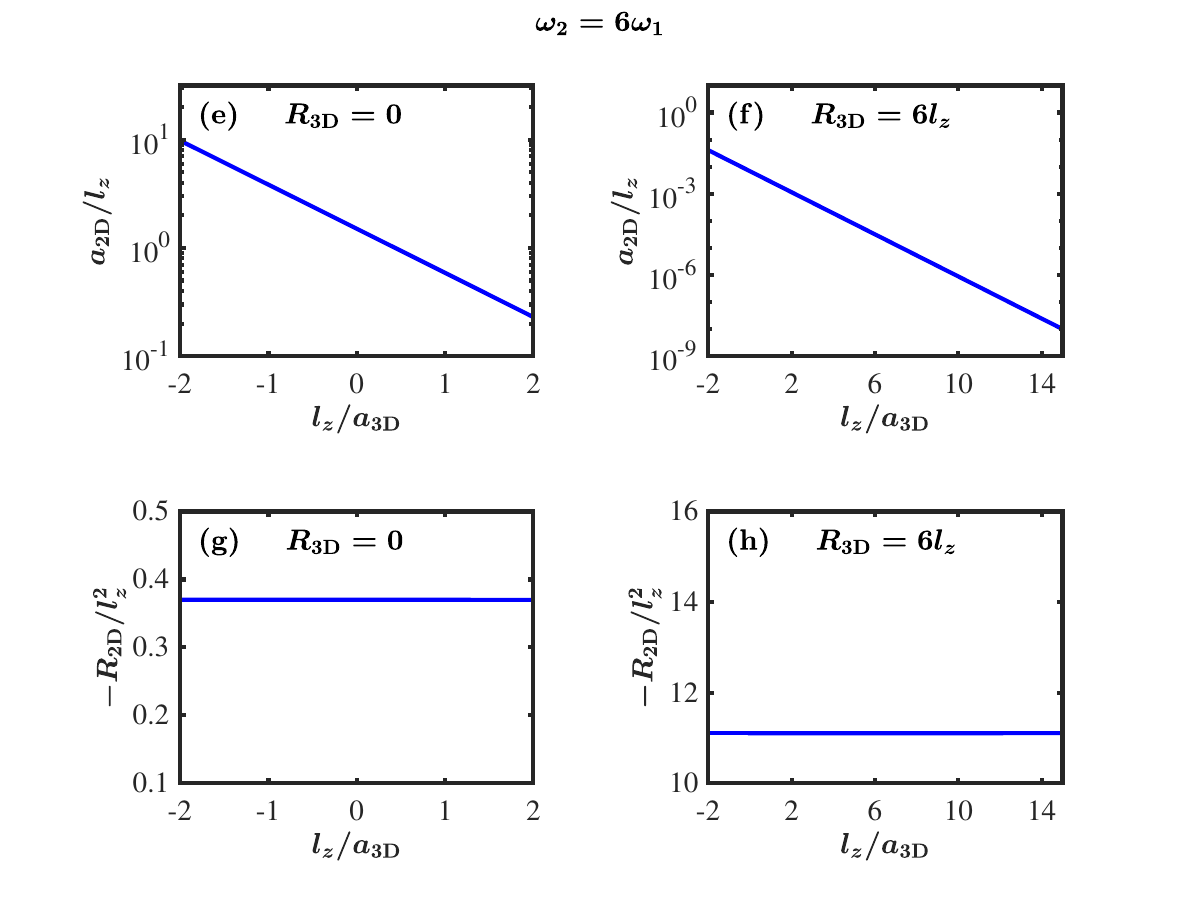}
        \caption{The 2D scattering length \(a_{\rm 2D}\) and effective range parameter
\(R_{\rm 2D}\) for a \(^{6}\mathrm{Li}\)-\(^{53}\mathrm{Cr}\) system
(\(m_1:m_2=6:53\)), in the case \(\omega_1<\omega_2\).
 In panels (a--d) and (e--h) show the results for
\(\omega_1:\omega_2=1:2\) and \(\omega_1:\omega_2=1:6\), respectively.
For both frequency ratios, the results are presented as functions of
\(\ell_z/a_{\rm 3D}\), with \(R_{\rm 3D}=0\) in panels (a,c,e,g) and
\(R_{\rm 3D}=6\ell_z\) in panels (b,d,f,h). Here
\(\ell_z=\sqrt{1/(m_1\omega_1)}\), as defined in Eq.~(\ref{l1mt}).
In panels (a,c,e,g). 
For details, see Sec.~\ref{2dsp}.
}
    \label{Fig3}
  \end{figure}

 \begin{figure}[t]
    \centering
    \includegraphics[width=1.0\linewidth]{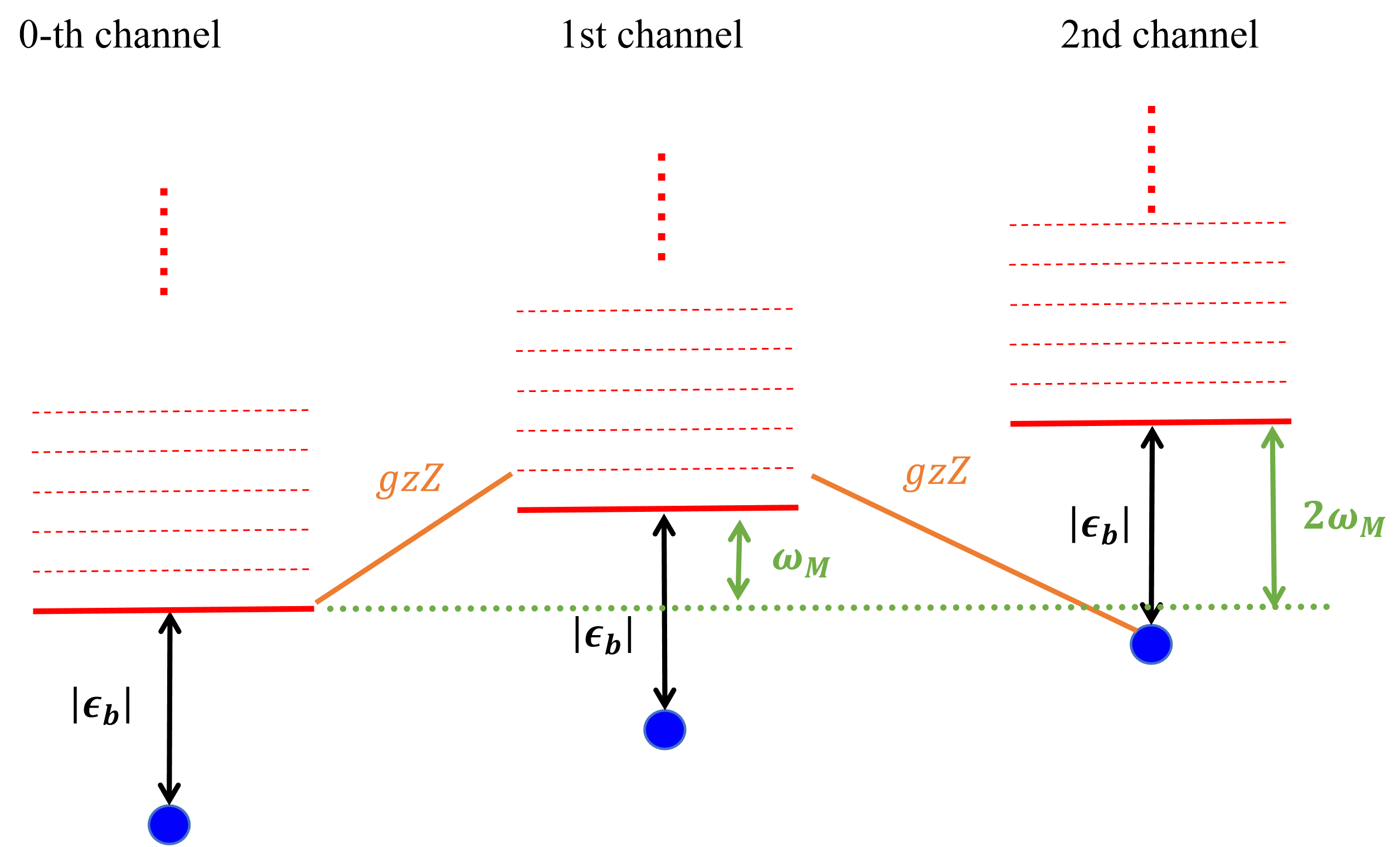}
  \caption{ Schematic illustration of the multichannel structure of the system. 
For clarity, only the lowest three CoM channels are shown. The solid red 
lines represent the threshold energies of the channels, the red dashed 
lines denote scattering states, and the blue dots indicate the bound 
states in the corresponding channels. The orange lines represent the 
inter-channel couplings induced by the term \(gzZ\). For details, see 
Sec.~\ref{2dsp}.
}
  \label{channel}
\end{figure}

In Fig.~\ref{Fig1}(a,c,e,g), we present the 2D scattering length
\(a_{\rm 2D}\) and the 2D effective range parameter \(R_{\rm 2D}\)
for the case \(\omega_1>\omega_2\), where the longitudinal confinement
frequency of the light atom is larger than that of the heavy atom.
Here the 3D effective range is set to zero, i.e., \(R_{\rm 3D}=0\).
For each frequency ratio, the results are shown as functions of the
3D scattering length \(a_{\rm 3D}\). We find that
\(a_{\rm 2D}\rightarrow \infty\) in the limit
\(a_{\rm 3D}\rightarrow 0^-\), as in the case with
\(\omega_1=\omega_2\) \cite{Petrov}. Moreover, multiple resonances
appear for \(a_{\rm 3D}>0\). In the vicinity of these resonances,
\(a_{\rm 2D}\) varies sharply from zero to infinity, while
\(|R_{\rm 2D}|\) is significantly enhanced.

The appearance of these resonances originates from the coupling between
the longitudinal CoM and relative motions, similar to
the case of a quasi-2D system with a longitudinal box confinement
potential~\cite{2024}. The
underlying mechanism can be understood as follows.
The Hamiltonian \(\hat H\) in Eq.~(\ref{h}) can be rewritten as
\begin{eqnarray}
\hat H=\hat H_\mu+\hat H_M+gzZ,\label{hag}
\end{eqnarray}
where
\begin{eqnarray}
\hat H_\mu&=&-\frac{1}{2\mu}\nabla_{\bm{r}}^2
+\frac{\mu \omega_\mu^2}{2}z^2+U(\bm{r}),\label{hmu}
\end{eqnarray}
is the Hamiltonian for the relative motion of the two atoms, with
\begin{eqnarray}
\omega_\mu=\sqrt{(m_2\omega_1^2+m_1\omega_2^2)/M},
\end{eqnarray}
and
\begin{eqnarray}
\hat H_M&=& -\frac{1}{2M}\frac{\partial^2}{\partial Z^2}
+\frac{M \omega_M^2}{2}Z^2\label{hm}
\end{eqnarray}
is the Hamiltonian for the longitudinal CoM motion, with
\begin{eqnarray}
\omega_M=\sqrt{(m_1\omega_1^2+m_2\omega_2^2)/M}.
\end{eqnarray}
Furthermore,
\begin{eqnarray}
g=2\mu\left(\omega_1^2-\omega_2^2\right)\label{gge}
\end{eqnarray}
is the strength of the coupling between the longitudinal CoM and
relative motions.

Eqs.~(\ref{hag})--(\ref{hm}) show that, as illustrated in
Fig.~\ref{channel}, the system can be viewed as a multichannel system.
The \(n\)th channel \((n=0,1,2,\ldots)\) corresponds to the \(n\)-th
eigenstate of the CoM Hamiltonian \(\hat H_M\). The relative motion in
each channel is governed by \(\hat H_\mu\), while the term \(gzZ\) in
Eq.~(\ref{hag}) couples different channels. 

Moreover, in the absence of the
inter-channel coupling \(gzZ\) , each channel contains a continuum of relative
scattering states and a single relative bound state, as shown in
Fig.~\ref{channel}. Specifically, the energy of the bound state in the
\(n\)-th channel is
\begin{eqnarray}
E_{bn}=E_{tn}-|\epsilon_b(a_{\rm 3D}, \omega_\mu,\mu)|,
\end{eqnarray}
where
\begin{eqnarray}
E_{tn}=(n+1/2)\omega_M+\omega_\mu/2
\end{eqnarray}
is the threshold energy of the \(n\)-th channel, and
\(|\epsilon_b(a_{\rm 3D}, \omega_\mu,\mu)|\) is the binding energy of
the bound state of the Hamiltonian \(\hat H_\mu\) in Eq.~(\ref{hmu}).
This binding energy depends on \(a_{\rm 3D}\), \(\omega_\mu\), and
\(\mu\) \cite{Petrov}.

When the inter-channel coupling \(gzZ\) is included, relative-motion
states in different channels become coupled. As a result, the coupled
system has only one scattering threshold, namely that of the 0th
channel. The parameters \(a_{\rm 2D}\) and \(R_{\rm 2D}\) characterize
the threshold scattering in this open channel in the presence of the
inter-channel coupling.

For convenience in the following discussion, we introduce the parities
\({\cal P}_z\) and \({\cal P}_Z\) with respect to the following two
transformations:
\begin{eqnarray}
{\cal P}_z&:& z_1\rightarrow z_2;\ \ \ \ \ \ z_1\rightarrow z_2
\ \ \ \ \  ({\rm i.e.,}\ z\rightarrow -z,\ Z\rightarrow Z),\nonumber\\
{\cal P}_Z&:& z_1\rightarrow -z_2;\ \ \ z_2\rightarrow -z_1
\ \ \ ({\rm i.e.,}\ z\rightarrow z,\ Z\rightarrow -Z).\nonumber
\end{eqnarray}
Since the coupling term \(gzZ\) changes both \({\cal P}_z\) and
\({\cal P}_Z\), it couples the threshold scattering state of the 0th
channel \(({\cal P}_z={\cal P}_Z=1)\) to the odd-wave scattering states
of the 1st channel \(({\cal P}_z={\cal P}_Z=-1)\), and further couples
the latter to the bound state of the 2nd channel
\(({\cal P}_z={\cal P}_Z=1)\), as illustrated in Fig.~\ref{channel}.
As a result, this bound state is effectively coupled to the threshold
scattering state of the 0th channel. Therefore, a resonance occurs
when these two states are resonant with each other, namely when
\(E_{b2}\approx E_{t0}\), or equivalently,
$
|\epsilon_b(a_{\rm 3D}, \omega_\mu,\mu)|\approx 2\omega_M.
$
The same argument shows that additional resonances appear when the
bound states of the 4th, 6th, and higher even channels become
resonant with the threshold scattering state of the 0th channel.
Specifically, our above analysis yield that the condition for the \(j\)th resonance
\((j=1,2,3,\ldots)\) can be estimated by the condition
\begin{eqnarray}
|\epsilon_b(a_{\rm 3D}, \omega_\mu,\mu)|=2j\omega_M.\label{con}
\end{eqnarray}
In Fig.~\ref{Fig1}(a,c,e,g), we mark the resonance positions estimated
from Eq.~(\ref{con}). These estimates agree very well with the exact
numerical results. 
 
 Moreover, in Fig.~\ref{Fig1}(b,d,f,h), we show \(a_{\rm 2D}\) and
\(R_{\rm 2D}\) for the same frequency ratios as those in
Fig.~\ref{Fig1}(a,c,e,g), but with a nonzero 3D effective range,
\(R_{\rm 3D}=6\ell_z\). We find that a nonzero \(R_{\rm 3D}\) shifts
the resonance positions and leads to a larger \(|R_{\rm 2D}|\). The
latter indicates a stronger energy dependence of the 2D scattering
amplitude, and can be attributed to the fact that the 3D
scattering amplitude itself has a stronger energy dependence for
\(R_{\rm 3D}\neq 0\) than for \(R_{\rm 3D}=0\).

In addition to the cases with \(\omega_1>\omega_2\) discussed above,
we also calculate the 2D scattering parameters \(a_{\rm 2D}\) and
\(R_{\rm 2D}\) for the opposite case, \(\omega_1<\omega_2\). The results
are shown in Fig.~\ref{Fig3}. 
No resonance is found in the parameter region shown in Fig.~\ref{Fig3}.
This observation is consistent with the above interpretation of the
resonance mechanism, since the estimate in Eq.~(\ref{con}) also predicts
the absence of resonances in this region. 
Eq.~(\ref{con}) shows that the first resonance occurs at \( l_z/a_{\rm 3D}\approx 2.43 \) for \(\omega_1:\omega_2 =1:2\) and at \( l_z/a_{\rm 3D}\approx 4.31 \) for \(\omega_1:\omega_2 =1:6\).
The
other behaviors are qualitatively similar to those for
\(\omega_1>\omega_2\), as shown in Fig.~\ref{Fig1}.

\subsection{Bound-State  Energy}
\label{sec:level3}

We now investigate the bound states. For the system described by the
Hamiltonian in Eq.~(\ref{h}), the total parity
\begin{eqnarray}
{\cal P}={\cal P}_z{\cal P}_Z,
\end{eqnarray}
which corresponds to the mirror reflection
\(\{z_1\rightarrow -z_1,z_2\rightarrow -z_2\}\), is conserved.
Therefore, the two subspaces with \({\cal P}=+1\) and \({\cal P}=-1\)
are decoupled. More specifically, the subspaces with \({\cal P}=+1\)
and \({\cal P}=-1\) are associated with the longitudinal basis states
\(\phi_m^{(1)}(z_1)\phi_n^{(2)}(z_2)\) with \(m+n\) even and odd,
respectively. In our system, when the two ultracold atoms are far apart,
they occupy the longitudinal ground state
\(\phi_0^{(1)}(z_1)\phi_0^{(2)}(z_2)\), which belongs to the
\({\cal P}=+1\) subspace. Therefore, only the states in this subspace
are relevant to the two-body  problem considered here \cite{space}. 

\begin{figure}[t]
    \centering
    \includegraphics[width=0.8\linewidth]{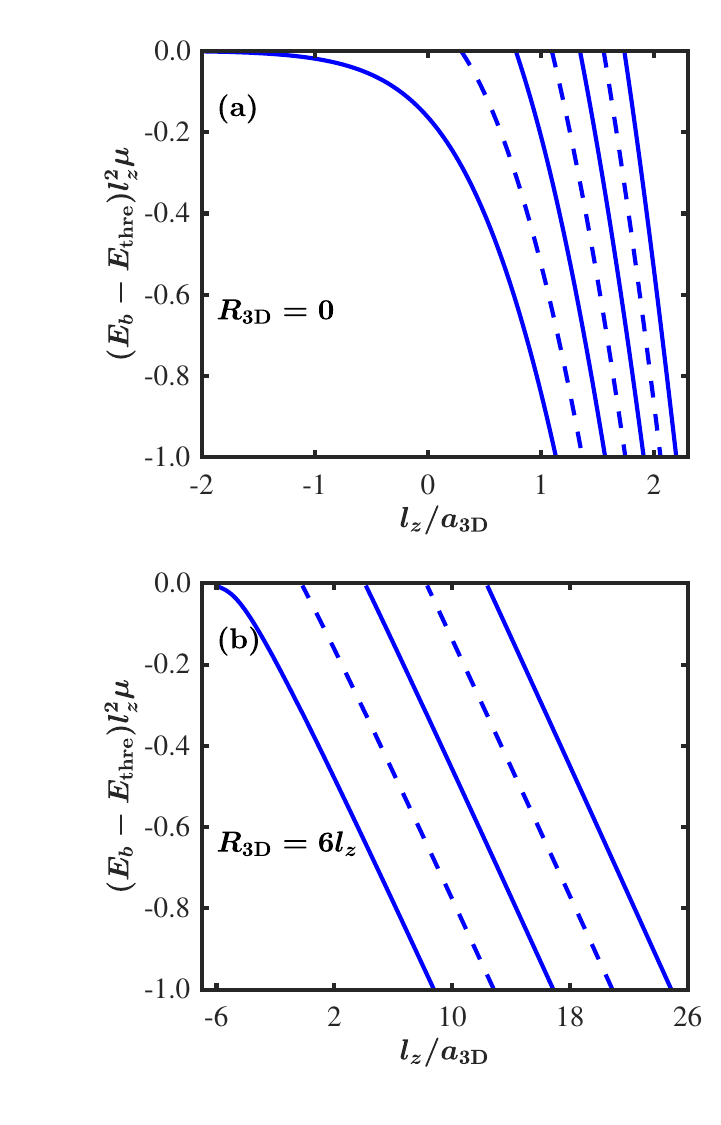}
  \caption{Bound-state energies \(E_b\) of a \(^{6}\mathrm{Li}\)-\(^{53}\mathrm{Cr}\)
system with \(m_1:m_2=6:53\) and
\(\omega_1:\omega_2=6:1\). Panels (a) and (b) show the
results for \(R_{\rm 3D}=0\) and \(R_{\rm 3D}=6\ell_z\), respectively.
Here \(E_{\rm thre}\equiv(\omega_1+\omega_2)/2\) is the scattering threshold,  corresponding to the longitudinal zero-point
energy. The solid and dashed
lines represent the bound states in the \({\cal P}=+1\) and
\({\cal P}=-1\) subspaces, respectively.}
  \label{Fig2}
\end{figure}

In Fig.~\ref{Fig2}, the solid lines
show the energies \(E_b\) of the bound states in the \({\cal P}=+1\)
subspace, obtained from our calculation. Specifically, we present the
results for \(\omega_1:\omega_2=6:1\), with
\(R_{\rm 3D}=0\) in Fig.~\ref{Fig2}(a) and
\(R_{\rm 3D}=6\ell_z\) in Fig.~\ref{Fig2}(b). In each case, there is
always one bound state whose energy approaches the longitudinal
zero-point energy
\(E_{\rm thre}\equiv(\omega_1+\omega_2)/2\), namely the scattering threshold, in
the limit \(a_{\rm 3D}\rightarrow 0^-\). In addition, several other
bound states are found, whose energies individually approach \(E_0\) at
the 2D resonance points shown in Fig.~\ref{Fig1}(a--d), i.e., the points where \(a_{\rm 2D}=\infty\). For comparison, Fig.~\ref{Fig2} also
shows the energies of the bound states in the \({\cal P}=-1\) subspace
as dashed lines.

\section{Summary}
\label{summary}

We exactly solve the two-body problem of two heteronuclear ultracold atoms
in a quasi-2D confinement, where the two atoms experience different
longitudinal confinement frequencies. The 2D scattering length
\(a_{\rm 2D}\), the effective range parameter \(R_{\rm 2D}\), and the two-body
bound-state energies are calculated, and multiple resonances induced
by the coupling between the longitudinal CoM and relative
motions are identified. Our calculation approach, which is summarized in
Appendix~\ref{appsum}, is applicable to ultracold atoms of arbitrary species.
In this paper, we use the \(^{6}\mathrm{Li}\)-\(^{53}\mathrm{Cr}\) system as a
representative example to quantitatively illustrate our results.

Our results can be directly used to manipulate the effective pairwise
interaction in quasi-2D heteronuclear ultracold gases, by tuning either the
3D scattering parameters or the longitudinal confinement frequencies. As
discussed above, such systems are hopeful to be experimentally
realized in the near future and may provide a useful platform for exploring
novel 2D many-body quantum phenomena.

\begin{acknowledgments}

This work is supported by the National Key Research and Development Program of China (Grant No.~2022YFA1405300), and  the Innovation Program for Quantum Science and Technology (Grant No.~2023ZD0300700). A paper by Tingting Shi and Xiaoling Cui, posted on arXiv today, independently addresses the same problem as ours. They solve the two-body problem exactly, using another approach. The approaches of the two works are equivalent, and the results are in agreement with each other.

\end{acknowledgments}


\begin{widetext}
\appendix

\section{Expression of the Operator $\hat A_{\rm 3D}$}
\label{a3d}

In this appendix we show the expression of the operator $\hat{A}_{\rm{3D}}$, which appears in the expression (\ref{uhy}) of the pseudo potential $U(\bm{r})$. $\hat{A}_{\rm{3D}}$ can be derived via the approach we developed in Ref.~\cite{LHY}. Here we just show the result. The mathematical symbols in this appendix, except for those newly defined, are defined the same as in the main text.

We first introduce the CoM longitudinal Hamiltonian $\hat{H}_{\mathrm{com}}$:
\begin{eqnarray}
\hat{H}_{\mathrm{com}}=-\frac{1}{2M}\frac{\partial^2}{\partial Z^2} + V^{(1)}(Z) + V^{(2)}(Z).
\end{eqnarray}
Moreover, we denote the eigen-energy and eigen-wavefunction of $\hat{H}_{\mathrm{com}}$ as $E_{\lambda}^{(\mathrm{com})}$ and $ \phi_{\mathrm{com}}^{(\lambda)}(Z)$, respectively ($\lambda=1,2,3,...$), i.e., we have
\begin{equation}\label{eq:com}
   \hat{H}_{\mathrm{com}} 
    \phi_{\mathrm{com}}^{(\lambda)}(Z) 
    = E_{\lambda}^{(\mathrm{com})} \phi_{\mathrm{com}}^{(\lambda)}(Z).
\end{equation}
Clearly, $\phi_{\mathrm{com}}^{(\lambda)}(Z)$ can also be denoted as a Dirac vector $\ket{\phi_{\mathrm{com}}^{(\lambda)}}$ in the Hilbert space of the longitudinal CoM motion.

The operator $\hat{A}_{\rm{3D}}$ can be expressed in terms of $\{E_{\lambda}^{(\mathrm{com})}\}$ and $\{\ket{\phi_{\mathrm{com}}^{(\lambda)}}\}$ as:
\begin{eqnarray}\label{eq:A3D}
    \frac{1}{\hat{A}_{\rm{3D}}(k)} &=& \sum_{\lambda}
    \left[ \frac{1}{a_{\rm{3D}}} + R_{\rm{3D}} \cdot 2\mu 
    \left( \frac{k^2}{2\mu} + \frac{\omega_1+\omega_2}{2} - E_{\lambda}^{(\mathrm{com})} \right) \right]
    \ket{\phi_{\mathrm{com}}^{(\lambda)}} \bra{\phi_{\mathrm{com}}^{(\lambda)}},\label{a3d1}\\
    &=& \frac{1}{a_{\rm{3D}}} + R_{\rm{3D}} \cdot 2\mu \left( \frac{\omega_1+\omega_2}{2} \right) 
       - R_{\rm{3D}} \cdot 2\mu \hat{H}_{\mathrm{com}} + R_{\rm{3D}} k^2.\label{a3d2}
\end{eqnarray}
Here the parameter $k$ has the following meanings: \\

\noindent {\bf (1) Scattering Problem:} For a scattering problem with incident momentum $\bm k$, we have $k=|\bm k|$, as in the main text. 

\noindent {\bf (2) Bound-State Problem:} In the calculation of a two-body bound state, we have $k=i\sqrt{2\mu |E_b|}$, where $E_b$ is the bound-state energy.\\

\noindent Furthermore, by defining  the operator $\hat{S}_0$ as
\begin{equation}
    \hat{S}_0 \equiv \frac{1}{a_{\rm{3D}}} + R_{\rm{3D}} \cdot 2\mu \left( \frac{\omega_1+\omega_2}{2} \right) 
           - R_{\rm{3D}} \cdot 2\mu \hat{H}_{\mathrm{com}},\label{s0}
\end{equation}
we can re-express $\hat{A}_{\rm{3D}}(k)$ as:
\begin{equation}
    \frac{1}{\hat{A}_{\rm{3D}}(k)} \equiv \hat{S}_0 + R_{\rm{3D}} k^2.\label{s0a3d}
\end{equation}

\section{Calculation of  2D Scattering Length  $a_{\rm 2D}$  and Effective Range Parameter $R_{\rm 2D}$}
\label{apps}

\subsection{Summary of the Main Steps}
\label{appsum}

In this appendix, we show our approach for the exact calculations of the 2D scattering length $a_{\rm{2D}}$ and effective range parameter $R_{\rm 2D}$, which is straightforwardly generalized from the one of our previous work in Ref.~\cite{2024}.
The main steps are summarized in this subsection, while  derivations are shown in subsequent sections.

We calculate $a_{\rm 2D}$ and $R_{\rm 2D}$ via the following steps:

{\bf Step 1}: Numerically solving the integral equations (\ref{ne1}, \ref{ne2}), and
derive the functions $\eta^{(0)}(Z)$ and $\eta^{(2)}(Z)$. All relevant terms in Eqs.~(\ref{ne1}, \ref{ne2}) are explicitly defined in the subsequent derivation.

{\bf Step 2}: Substituting the solutions $\eta^{(0,2)}(Z)$, obtained in  Step 1, into Eqs. (\ref{a2deff}) and (\ref{reff}) to evaluate the scattering parameters $a_{\rm 2D}$ and $R_{\rm 2D}$.

The following subsections present detailed derivations of Eqs.~(\ref{ne1}, \ref{ne2}, \ref{a2deff}, \ref{reff}) and demonstrate the validity of the above steps.
The mathematical symbols in this appendix, except for those newly defined, are defined the same as in the main text and Appendix~\ref{a3d}.

\subsection{Lippmann-Schwinger Equation }

As shown in our main text, we study the scattering of two heteronuclear ultracold atoms 
1 and 2 in a species-dependent quasi-2D harmonic confinement, i.e.,  the system configuration detailed in Sec.~\ref{sec:level2}. The incident scattering state $\Psi_{\bm k}^{\rm(in)}({\bm \rho},z_1,z_2)$ is given by Eq.~(\ref{incident}). Since the interatomic interaction is modeled by the zero-range  pseudopotential $U(\bm{r})$ of Eq.~(\ref{uhy}), the scattering process is restricted to the $L_z=0$ angular momentum subspace, with $L_z$ representing the $z$-component of the relative orbital angular momentum. Consequently, we focus on the projection of the scattering wave function $\Psi_{\bm k}^{(+)}({\bm \rho},z_1,z_2)$ of Eq.~(\ref{psiexp0}) in this subspace, which is denoted as $\Psi_k^{(s)}(\rho, z_1, z_2)$. Here $k=|{\bm k}|$ and $\rho=|{\bm \rho}|$ are the magnitudes of the in-plane momentum and relative coordinate, respectively. Clearly, $\Psi_{\bm k}^{(+)}({\bm \rho},z_1,z_2)$ can be expressed as:
\begin{eqnarray}
\Psi_k^{(s)}(\rho, z_1, z_2)= \sum_{m, n=0}^{\infty}\phi_m^{(1)}(z_1) \phi_n^{(2)}\left(z_2\right)\Psi_{k}^{(m,n)}(\rho),\label{psiexp}
\end{eqnarray}
where $\Psi_{ k}^{(m,n)}(\rho)$ is the projection of $\psi_{\bm k}^{(m,n)}({\bm \rho})$ of Eq.(\ref{psiexp0}) in the subspace with $L_z=0$.
Additionally, $\Psi_k^{(s)}(\rho, z_1, z_2)$ satisfies the Schr\"odinger equation
\begin{eqnarray}
{\hat H}\Psi_k^{(s)}\left(\rho, z_1, z_2\right)=E\Psi_k^{(s)}\left(\rho, z_1, z_2\right),\ \ \ {\rm with}\ \
E=\frac{k^2}{2\mu}+\frac{\omega_1}{2}+\frac{\omega_2}{2}<\frac{\omega_{1}+\omega_{2}}{2}+\omega_{1,2},
\end{eqnarray}
as well as the  conditions
\begin{eqnarray}
\lim_{\rho\rightarrow\infty}\Psi_{k}^{(m,n)}(\rho)&=&0,\ \ \  \  {\rm for} \ \ (m,n)\neq (0,0),
\\[5pt]
\Psi_{k}^{(0,0)}(\rho)&=&J_0(k\rho) + \frac{-1}{i\frac{\pi}{2}-\gamma- \ln\frac{ka_{\rm{2D}}}{2}-\frac{1}{2}R_{\rm 2D}k^2}K_0(-ik\rho), \ \ \ \   {\rm for} \ \rho>0,
\label{conapp1}
\end{eqnarray}
which are lead by the conditions (\ref{con-1}, \ref{con0}).
Here
$J_0$ and $K_0$ are the Bessel function of the first kind and the modified Bessel function of the second kind, respectively.
We emphasize that Eq.~(\ref{conapp1}) is exact for all $\rho>0$, as a direct consequence of the zero-range character of the  pseudopotential $U(\bm{r})$.

Moreover, we introduce another wave function $\Phi_k\left(\rho, z_1, z_2\right)$ for the convenience of the further discussions, which satisfies the stationary Schr\"odinger equation
\begin{eqnarray}
{\hat H}\Phi_k\left(\rho, z_1, z_2\right)=E\Phi_k\left(\rho, z_1, z_2\right),
\end{eqnarray}
together with  the long-range boundary condition
\begin{eqnarray}
\lim_{\rho\rightarrow\infty}\Phi_{k}^{(m,n)}(\rho)&=&0,\ \ \ \ \ \ \ \ \ \ \ \ \ \ \  \ \ \ \ \ \ \ \ \ \  \   {\rm for} \ \ (m,n)\neq (0,0),\label{conapp0}\\
\Phi_{k}^{(0,0)}(\rho)&=&J_0(k\rho)+{\cal C}X_0(k,\rho),\ \ \ {\rm for}\ \rho>0. \label{conapp}
\end{eqnarray}
Here  the function $X_0(k,\rho)$ is defined as
\begin{eqnarray}
X_0(k, \rho)= K_0(-ik\rho) + \left[-\frac{i\pi}{2}+\gamma+\ln(k\ell_z)\right] J_0(k\rho),
\end{eqnarray}
with 
\begin{eqnarray}
\ell_z= \sqrt{\frac{1}{m_1\omega_1}},\label{l1}
\end{eqnarray}
and ${\cal C}$ is a to-be-determined constant. We can expand $\Phi_k(\rho, z_1, z_2)$ in terms of the longitudinal wave functions $\phi_m^{(1)}(z_1) \phi_n^{(2)}\left( z_2\right)$ as
\begin{eqnarray}
\Phi_k(\rho, z_1, z_2)=\sum_{m, n=0}^{\infty}\phi_m^{(1)}(z_1) \phi_n^{(2)}\left( z_2\right)\Phi_{k}^{(m,n)}(\rho).\label{phiexp}
\end{eqnarray}
The explicit meaning of the condition (\ref{conapp}) can be understood with this expansion: when $\Phi_{k}^{(0,0)}(\rho)$ of Eq.~(\ref{phiexp}) is expressed as  a linear  combination of the functions $J_0(k\rho)$ and $X_0(k, \rho)$ at limit $\rho\rightarrow\infty$, then the  coefficient of $J_0(k\rho)$ is normalized to unity.

Both $\Phi_k\left(\rho, z_1, z_2\right)$ and the  scattering wave function   $\Psi_k^{(s)}\left(\rho, z_1, z_2\right)$ with $L_z=0$ satisfy the same stationary Schr\"odinger equation and the same
long-range boundary condition for the excited  longitudinal  modes,
but differ in the ground  longitudinal  mode. Consequently, they are proportional:
\begin{eqnarray}
\Phi_k\left(\rho, z_1, z_2\right) \propto \Psi_k^{(s)}\left(\rho, z_1, z_2\right).\label{prop}
\end{eqnarray}
Since the scattering parameters $a_{\rm 2D}$ and $R_{\rm 2D}$ are determined by the behavior of $\Psi_k^{(s)}\left(\rho, z_1, z_2\right)$, they can equivalently be extracted from  $\Phi_k\left(\rho, z_1, z_2\right)$. Thus, in the following we focus on
$\Phi_k$. Specifically, in this and the following subsection, we analyze the Lippmann-Schwinger equation (LSE) for  $\Phi_k\left(\rho, z_1, z_2\right)$, and introduce the auxiliary functions $\eta^{(0,2)}(Z')$ from this LSE.
In Sec. \ref{sss} we  derive $a_{\rm 2D}$ and $R_{\rm 2D}$ from $\Phi_k\left(\rho, z_1, z_2\right)$.

The LSE of  $\Phi_k\left(\rho, z_1, z_2\right)$ can be expressed as
 ($\hbar=1$):
\begin{align}
  \Phi_k\left(\rho, z_1, z_2\right)=\phi_0^{(1)}(z_1) \phi_0^{(2)}(z_2)J_0(k\rho)+ \int_{-\infty}^{+\infty}dz_1'\int_{-\infty}^{+\infty}dz_2'\int_0^{\infty}d\rho^\prime
  G_E\bigg(\rho, z_1, z_2 ; \rho^\prime, z_1^{\prime}, z_2^{\prime}\bigg) D_k\left(\rho^\prime, z_1^{\prime}, z_2^{\prime}\right).\label{eq.LS-equation}
\end{align}
and the function $D_k\left(\rho^\prime, {z_1^\prime, z_2^\prime}\right)$ is defined as
\begin{equation}
 D_k\left(\rho^\prime, z_1^\prime,z_2^\prime\right) ={\frac{2 \hat{A}_{\rm{3D}}(k)}{\mu}}
 \delta(z_1^\prime-z_2^\prime)\delta(\rho^\prime)
  \frac{1}{\rho^\prime}\frac{\partial}{\partial \rho^\prime}\bigg[\rho^\prime\Phi_k\big({\rho^\prime},z_1^\prime,z_2^\prime\big)\bigg],
\end{equation}
here $\hat{A}_{\rm{3D}}(k)$ is the operator given by Eqs.~(\ref{a3d1}, \ref{a3d2}). Additionally, in Eq.~(\ref{eq.LS-equation})
$G_E$ is the free Green's function  in the subspace $L_z=0$, 
corresponding to the boundary conditions (\ref{conapp0}, \ref{conapp}) of $\Phi_k$,
and can be expressed as:
\begin{eqnarray}
  G_E\bigg(\rho, z_1, z_2 ; \rho^\prime, z_1^{\prime}, z_2^{\prime}\bigg)= \sum_{m, n=0}^{\infty}\phi_m^{(1)}(z_1) \phi_m^{(1)}\left(z_1^{\prime}\right)\phi_n^{(2)}(z_2)\phi_n^{(2)}\left(z_2^{\prime}\right) g_{2D}^{(\frac{k^2}{2\mu}-m\omega_1-n\omega_2)}({\rho},{\rho}^{\prime}),\label{eq.G4d}
\end{eqnarray}
where
\begin{eqnarray}
 g_{2D}^{({\cal E})}(\rho,\rho') &=&
 \left\{
    \begin{array}{cc}
      \tilde{\alpha}(\xi,\rho') J_0(\xi\rho) & (\rho<\rho') \\
      \\
      \tilde{\beta}(\xi,\rho') X_0(\xi,\rho) & (\rho>\rho')
    \end{array}\right., \hspace{1cm} ({\rm for}\ {\cal E}\geq 0);\\
    \nonumber\\
        \nonumber\\
    g_{2D}^{({\cal E})}(\rho,\rho') &=&
    \left\{
    \begin{array}{cc}
      \alpha(\kappa,\rho') I_0(\kappa\rho) & (\rho<\rho')\\
      \\
      \beta(\kappa,\rho') K_0(\kappa\rho) & (\rho>\rho')
    \end{array}\right., \hspace{1.2cm} ({\rm for}\ {\cal E}< 0),
\end{eqnarray}
with
\begin{eqnarray}
\xi=\sqrt{2\mu{\cal E}},\ \ \ \ \kappa=\sqrt{2\mu|{\cal E}|}.
\end{eqnarray}
Here $I_j$ and $K_j$  ($j=0,1,2,...$) denote the modified Bessel functions of the first and second kind, respectively,
and $\tilde{\alpha}$, $\tilde{\beta}$, $\alpha$ and
$\beta$ are defined as
\begin{eqnarray}
  \tilde{\alpha}(\xi,\rho') &=&\frac{2\mu X_0(\xi,\rho')}{J_0(\xi\rho')X_0^\prime(\xi,\rho')-J_0^\prime(\xi\rho')X_0(\xi,\rho')},\ \ \
  \tilde{\beta}(\xi,\rho') =\frac{2\mu J_0(\xi\rho')}{J_0(\xi\rho')X_0^\prime(\xi,\rho')-J_0^\prime(\xi\rho')X_0(\xi,\rho')},\\[5pt]
   \alpha(\kappa,\rho') &=& \frac{2\mu K_0(\kappa\rho')}{\kappa\left[K_0(\kappa\rho')I_1(\kappa\rho')-K_1(\kappa\rho')I_0(\kappa\rho')\right]},\
    \beta(\kappa,\rho') = \frac{2\mu I_0(\kappa\rho')}{\kappa\left[K_0(\kappa\rho')I_1(\kappa\rho')-K_1(\kappa\rho')I_0(\kappa\rho')\right]},\\
    \nonumber
\end{eqnarray}
with
$J_0^\prime(\xi\rho')=\frac{d }{d\rho'}{J_0(\xi\rho')}$ and $X_0^\prime(\xi\rho')=\frac{d }{d\rho'}{X_0(\xi\rho')}$.

\subsection{Definition of the Functions  $\eta^{(0,2)}(Z')$}

We now re-express Eq.~(\ref{eq.LS-equation})  as
\begin{eqnarray}
  \Phi_k(\rho, z_1, z_2)=\phi_0^{(1)}(z_1)\phi_0^{(2)}(z_2)J_0(k\rho)+ \frac{1}{2}\int_{-\infty}^{+\infty}\mathrm{d}Z'\Omega_k\left(\rho,z_1, z_2,Z'\right)\eta_k(Z'),\label{lser}
\end{eqnarray}
%
%
where the functions $\eta_k(Z)$ and $\Omega_k(\rho,z_1, z_2,Z') $ are defined by
  \begin{eqnarray}
  \eta_k(Z)=\ \lim_{\rho\to 0}\frac{2\hat{A}_{\rm{3D}}(k)}{\mu}\frac{\partial}{\partial \rho}\left[\rho\cdot\Phi_k(\rho,Z,Z)\right],\label{eta}
\end{eqnarray}
and
\begin{eqnarray}
  \Omega_k(\rho,z_1,z_2,Z')
  =
  \sum_{m,n=0}^{\infty}\phi_m^{(1)}(z_1)\phi_n^{(2)}(z_2)\phi_m^{(1)}(Z^{\prime})\phi_n^{(2)}(Z^{\prime})\cdot\Lambda_{\rm{2D}}^{(\frac{k^2}{2\mu}-m\omega_1-n\omega_2)}(\rho),\label{eq.omega}
\end{eqnarray}
respectively. Here the function $\Lambda_{\rm{2D}}^{({\cal E})}(\rho)$ is defined as
  \begin{eqnarray}
    \Lambda_{\rm{2D}}^{({\cal E})}(\rho) = \left\{
    \begin{array}{ll}
      -2\mu X_0(\sqrt{2\mu{\cal E}}, \rho),
      &\hspace{1cm} ({\cal E}\geq 0) \\
      \\
      -2\mu K_0(\sqrt{2\mu|{\cal E}|}\rho), &\hspace{1cm}({\cal E}< 0)\
    \end{array}\right..\label{eq.x0}
  \end{eqnarray}
In the short-range regime $\rho\ll 1/k$, we can further expand both the scattering wave function $\Phi_k(\rho, z_1, z_2)$ and the auxiliary function $\eta_k(Z') $ as
\begin{eqnarray}
\Phi_k(\rho, z_1, z_2) = \Phi^{(0)}(\rho, z_1, z_2) + k^2 \Phi^{(2)}(\rho, z_1, z_2)+ \mathcal{O}(k^4),\label{psiexp}
\end{eqnarray}
and
\begin{eqnarray}
    \eta_k(Z') = \eta^{(0)}(Z') + k^2\eta^{(2)}(Z') + \mathcal{O}(k^4),
\end{eqnarray}
respectively. 
Substituting the above expansions into Eq.~(\ref{eta}) and using Eq.~(\ref{s0a3d}), we find that $\eta^{(j)}(Z')$ $(j=0,2)$ are related to $\Phi^{(j)}(\rho, z_1,z_2)$ through the operator $\hat{S_0}$ defined in Eq.~(\ref{s0}):
\begin{eqnarray}\label{eq.eta}
	\hat{S_0}[\eta^{(0)} (Z')]&=&\frac{2
		}{\mu}\frac{\partial}{\partial \rho'}\left[\rho'\cdot \Phi^{(0)}(\rho',Z',Z')\right] \bigg|_{\rho' = 0},
	   \\
	\hat{S_0}[\eta^{(2)} (Z')]+R_{\rm3D}\eta^{(0)} (Z')&=&\frac{2
		}{\mu}\frac{\partial}{\partial \rho'}\left[\rho'\cdot\Phi^{(2)}(\rho',Z',Z')\right]\bigg|_{\rho' = 0}.
\end{eqnarray}
Moreover, expanding both sides of Eq.~(\ref{lser})  up to oder $k^2$ in the region with $\rho\ll 1/k$,
we obtain
\begin{eqnarray}
  \Phi^{(0)}(\rho, z_1, z_2)&=&\phi_0^{(1)}(z_1)\phi_0^{(2)}(z_2)+ \frac{1}{2}\int_{-\infty}^{+\infty}\Omega_{k=0}(\rho,z_1,z_2,Z')\eta^{(0)}(Z')\mathrm{d}Z',\label{eq.LS-omega0}\\
  \nonumber\\
  \Phi^{(2)}(\rho, z_1, z_2)&=&-\phi_0^{(1)}(z_1)\phi_0^{(2)}(z_2)\frac{\rho^2}{4}+ \frac{1}{2}\int_{-\infty}^{+\infty}\Omega^{(2)}(\rho,z_1,z_2,Z')\eta^{(0)}(Z')\mathrm{d}Z'+ \frac{1}{2}\int_{-\infty}^{+\infty}\Omega_{k=0}(\rho,z_1,z_2,Z')\eta^{(2)}(Z')\mathrm{d}Z',\nonumber\\
  \label{eq.LS-omega2}
\end{eqnarray}
where the function $\Omega^{(2)}(\rho,z_1,z_2,Z')$ is defined as
\begin{eqnarray}
\Omega^{(2)}(\rho,z_1,z_2,Z') &=&\sum_{m,n=0}^{\infty}\phi_m^{(1)}(z_1)\phi_n^{(2)}(z_2)\phi_m^{(1)}(Z')\phi_n^{(2)}(Z')\Lambda_{2}^{(mn)}(\rho),\label{omega2}
\end{eqnarray}
with
\begin{eqnarray}
  \Lambda_2^{(mn)}(\rho) =
  \left\{
  \begin{array}{ll}
    -\dfrac{\mu}{2}\rho^2\ln\left(\dfrac{\rho}{2\ell_z}\right), & (m=n=0)\\
    \\
    \sqrt{\dfrac{\mu}{2|m\omega_1+n\omega_2|}}\rho K_1\left[\sqrt{2\mu(m\omega_1+n\omega_2)}\rho\right], & ((m,n)\neq (0,0))
  \end{array}\right..
\end{eqnarray}

\subsection{Derivation of Eqs.~(\ref{a2deff}, \ref{reff})}
\label{sss}

Substituting Eqs.~(\ref{eq.LS-omega0}, \ref{eq.LS-omega2}) into Eq.~(\ref{psiexp}) and then into Eq.~(\ref{phiexp}), we obtain the short-range expansion of  $\Phi_{k}^{(0,0)}(\rho)$:
\begin{eqnarray}
\Phi_{k}^{(0,0)}(\rho)=1 + \mu\ln\left(\frac{\rho}{2\ell_z}\right)\left\{\int_{-\infty}^{+\infty}\mathrm{d}Z'
\phi_0^{(1)}(Z')\phi_0^{(2)}(Z')
\left[\eta^{(0)}(Z')+k^2\eta^{(2)}(Z')\right]\right\}+\mathcal{O}(\rho)+\mathcal{O}(k^4),\ \ 
 {\rm for}\ \rho\ll 1/k.\nonumber\\
\label{psik11}
\end{eqnarray}
On the other hand, according to the condition(\ref{conapp1}), the ground  longitudinal component  $\Psi_{k}^{(0,0)}(\rho)$ of the scattering wave function $\Psi_k^{(s)}\left(\rho, z_1, z_2\right)$  with $L_z=0$ satisfies
\begin{eqnarray}
  \Psi_{k}^{(0,0)}\left(\rho, z_1, z_2\right) \propto \ln(\rho/a_{\rm{2D}}) -\frac{1}{2}R_{\rm 2D}k^2 +\mathcal{O}(\rho)+\mathcal{O}(k^4),\ \ \ \ {\rm for}\ \rho\ll 1/k.\label{eq.pure-2d-asym}
\end{eqnarray}
Furthermore, the proportionality $\Phi_k \propto \Psi_k^{(s)}$ from Eq.~(\ref{prop}) implies that their  longitudinal-ground-state projections also satisfy $\Phi_{k}^{(0,0)}(\rho) \propto \Psi_{k}^{(0,0)}(\rho, z_1, z_2)$. This relation allows us to match the asymptotic behaviors, and solving the system formed by Eqs.~(\ref{psik11}) and (\ref{eq.pure-2d-asym}) yields
\begin{eqnarray}
&&1 + \mu\ln\left(\frac{\rho}{2\ell_z}\right)\left\{\int_{-\infty}^{+\infty}\mathrm{d}Z'
\phi_0^{(1)}(Z')\phi_0^{(2)}(Z')
\left[\eta^{(0)}(Z')+k^2\eta^{(2)}(Z')\right]\right\}\nonumber\\[5pt]
&=&
\mu\left\{\int_{-\infty}^{+\infty}\mathrm{d}Z'
\phi_0^{(1)}(Z')\phi_0^{(2)}(Z')
\left[\eta^{(0)}(Z')+k^2\eta^{(2)}(Z')\right]\right\}\left[
\ln(\rho/a_{\rm{2D}}) -\frac{1}{2}R_{\rm 2D}k^2\right],
\end{eqnarray}
which gives
\begin{eqnarray}
 a_{\rm{2D}}&=&\ 2\ell_z\exp\left[\frac{-1}{\mu \int\phi_0^{(1)}(Z')\phi_0^{(2)}(Z')\eta^{(0)}(Z')\mathrm{d}Z'}\right],\label{a2deff}\\
  \nonumber\\[5pt]
  R_{\rm 2D}&=&\frac{2}{\mu}\frac{\int\phi_0^{(1)}(Z')\phi_0^{(2)}(Z')\eta^{(2)}(Z')\mathrm{d}Z'}{\left[ \int\phi_0^{(1)}(Z')\phi_0^{(2)}(Z')\eta^{(0)}(Z')\mathrm{d}Z'\right]^2}.\label{reff}
\end{eqnarray}



\subsection{Derivations of Eqs.~(\ref{ne1}, \ref{ne2})}
\label{sec.app-b}


By combining Eq.~(\ref{eta}) with the Bethe–Peierls boundary condition obeyed by the scattering state $\Phi_k(\rho,z_1,z_2)$, which originates from the use of the  pseudopotential $U(\bm{r})$ of Eq.~(\ref{uhy}), and using Eq.~(\ref{s0a3d}), we obtain
\begin{eqnarray}
 \Phi_k(\rho,Z,Z)&=&-\frac{\mu}{2}\left(\frac{1}{\rho}\eta(Z) -\frac{1}{\hat{A}_{\rm{3D}}}\eta(Z) \right) + \mathcal{O}(\rho)\nonumber \\[5pt]
 &=&-\frac{\mu}{2}\left(\frac{1}{\rho}\eta(Z) -\hat{S_0}[\eta(Z)]-R_{\rm3D}  k^2 \eta(Z)\right) + \mathcal{O}(\rho),\label{eqeta2}
\end{eqnarray}
where $\hat{S_0}$ is defined in Eq.~(\ref{s0}).
Eq.~(\ref{eqeta2}) allows one to formally express the quantities  $\eta^{(0,2)}(Z)$ as
\begin{eqnarray}
  \frac{\mu}{2}\hat{S_0} \big[\eta^{(0)}(Z)\big]&=&\hat{\mathbb O}^{(0)}(Z)\bigg[\Phi^{(0)}(\rho,Z,Z)\bigg],\label{eq.eqhatO}
  \\
   \frac{\mu}{2}\hat{S_0}\big[ \eta^{(2)}(Z)\big]&=&\hat{\mathbb O}^{(2)}(Z)\bigg[\Phi^{(2)}(\rho,Z,Z)\bigg]-\frac{\mu}{2}R_{\rm3D} \eta^{(0)}(Z).\label{eq.eqhat2}
\end{eqnarray}
Here  $\hat{\mathbb O}^{(j)}(Z)$ is defined as follows: for any function $f(\rho,Z)$, we have
\begin{eqnarray}
\hat{\mathbb O}^{(j)}(Z)\bigg[f(\rho,Z)\bigg]\equiv\lim_{\rho\to 0}\left[f(\rho,Z)+ \frac{\mu}{2\rho}\eta^{(j)}(Z)\right],\ \ \ \ (j=0,2).
\end{eqnarray}
Subsequently, applying $\hat{\mathbb O}(Z)$ to both sides of Eqs.~(\ref{eq.LS-omega0}) and (\ref{eq.LS-omega2}) with $z_1=z_2=Z$, we derive coupled integral equations for  $\eta^{(0,2)}(Z)$ as
\begin{eqnarray}
  \frac{\mu}{2}\hat{S_0}\left[\eta^{(0)}(Z)\right]&=&\phi_0^{(1)}(Z)\phi_0^{(2)}(Z)+ \frac{1}{2}\lim_{\rho\to 0}\mathcal{I}^{(0)}(\rho,Z)\label{eq.eta0}\\
  \nonumber\\
 \frac{\mu}{2}\hat{S_0}\left[\eta^{(2)}(Z)\right]+\frac{\mu}{2}R_{\rm3D} \eta^{(0)}(Z)&=& \frac{1}{2}\int_{-\infty}^{+\infty}\Omega^{(2)}(0,Z,Z,Z')\eta^{(0)}(Z')\mathrm{d}Z'+ \frac{1}{2}\lim_{\rho\to 0}\mathcal{I}^{(2)}(\rho,Z),\label{eq.eta2}
\end{eqnarray}
with:
\begin{eqnarray}
  \mathcal{I}^{(j)}(\rho,Z)&\equiv&\int_{-\infty}^{+\infty}\Omega_{k=0}(\rho,Z,Z,Z')\eta^{(j)}(Z')\mathrm{d}Z' + \frac{\mu}{\rho}\eta^{(j)}(Z),\ \ \ \ (j=0,2).\label{eq.I}
\end{eqnarray}
We can further re-express the terms $\lim_{\rho\to 0}\mathcal{I}^{(0,2)}(\rho,Z)$ of Eqs. (\ref{eq.eta0}, \ref{eq.eta2}) as Hadamard finite part integral. Specifically, using the fact
\begin{eqnarray}
  \frac{1}{\rho} &=& \frac{1}{\pi}\int_{-\infty}^{+\infty}\frac{b}{\rho^2+b^2(z-z^{\prime})^2}\mathrm{d}z^{\prime}, \ \ \ \forall b>0,
      \end{eqnarray}
we can obtain
\begin{eqnarray}
   \lim_{\rho\to 0}\mathcal{I}^{(j)}(\rho,Z)&=&\lim_{\rho\to 0}\left[\int_{-\infty}^{+\infty}\Omega_{k=0}(\rho,Z,Z,Z')\eta^{(j)}(Z')\mathrm{d}Z' + \int_{-\infty}^{+\infty}\frac{\mu}{\pi}\frac{b\eta^{(j)}(Z)}{\rho^2+b^2(Z-Z^{\prime})^2}\mathrm{d}Z'\right]\nonumber\\
  &=&\lim_{\rho\to 0}\lim_{\epsilon\to 0}\left[\int_{-\infty}^{Z'-\epsilon}\Omega_{k=0}(\rho,Z,Z,Z')\eta^{(j)}(Z')\mathrm{d}Z'+\int^{+\infty}_{Z^{\prime}+\epsilon}\Omega_{k=0}(\rho,Z,Z,Z')\eta^{(j)}(Z')\mathrm{d}Z'\right.\nonumber\\
  &&\ \ \ \ \ \ \ \ \ \ \left.+ \int_{-\infty}^{Z'-\epsilon}\frac{\mu}{\pi}\frac{b\eta^{(j)}(Z)}{\rho^2+b^2(Z-Z^{\prime})^2}\mathrm{d}Z' + \int_{Z'+\epsilon}^{+\infty}\frac{\mu}{\pi}\frac{b\eta^{(j)}(Z)}{\rho^2+b^2(Z-Z^{\prime})^2}\mathrm{d}Z'\right]. \label{ehad1}
    \end{eqnarray}
Notice that Eq.~(\ref{ehad1}) holds for any positive value of $b$.
To regularize the divergent integral, we introduce a specific parameter $b_0>0$, defined such that when $b=b_0$, the integral in Eq.~(\ref{ehad1}) becomes uniformly convergent in the  limit $\rho\rightarrow 0$, Under this condition, the limiting operation 
lim $\lim_{\rho\rightarrow 0}$ can be interchanged with the integration.
The value of  $b_0$ will later be determined from the short-distance behavior of  $\Omega_{k=0}(\rho,Z,Z,Z')$ as $\rho\to 0$.
Taking $b=b_0$,  Eq.~(\ref{ehad1}) can be written as
\begin{eqnarray}
  \lim_{\rho\to 0}\mathcal{I}^{(j)}(\rho,Z)  &=& \mathbb{Z}\int \Omega_{k=0}(\rho,Z,Z,Z')\eta^{(j)}(Z')\mathrm{d}Z',\nonumber\\
  &\equiv&\lim_{\epsilon\to 0}\left[\int_{-\infty}^{Z'-\epsilon} \Omega_{k=0}(0,Z,Z,Z')\eta^{(j)}(Z')\mathrm{d}Z' + \int_{Z'+\epsilon}^{+\infty}\Omega_{k=0}(0,Z,Z,Z')\eta^{(j)}(Z')\mathrm{d}Z' + \frac{2}{b_0}\frac{\mu}{\pi}\frac{\eta^{(j)}(Z)}{\epsilon} \right],\nonumber\\
  &&\hspace{12cm} (j=0,2).
\label{ehad2}
\end{eqnarray}
Here $\mathbb{Z}\int\cdots\mathrm{d}Z^{\prime}$ is called as the Hadamard finite part integral. It should be noted that the Hadamard finite part integral in Eq.~(\ref{ehad2}) depends solely on the off-diagonal contributions of  $\Omega_{k=0}(0,Z,Z,Z')$, {\it i.e.}, the values $Z\neq Z'$.

To compute  $\Omega_{k=0}(0,Z,Z,Z')$ and derive the value of $b_0$, we utilize the Laplace representation{
\begin{eqnarray}
 - 2\mu K_0\left(\sqrt{2\mu|\mathcal{E}|}\rho\right)&=&-\int_0^{+\infty}{\mathrm d}\tau \frac{\mu}{\tau}e^{\tau\mathcal{E}-\frac{\mu\rho^2}{2\tau}},\ \ \ \ \ \ \ (\mathcal{E}<0).\label{eq.laplace-k}
\end{eqnarray}}
to re-express  Eq.~(\ref{eq.omega}) as
\begin{eqnarray}
  \Omega_{k}(\rho,Z,Z,Z')&=&-2\mu X_0(k,\rho)\phi_0^{(1)}(Z)\phi_0^{(2)}(Z)\phi_0^{(1)}(Z')\phi_0^{(2)}(Z')\nonumber\\
&&  -\sum_{(m,n)\neq(0,0)}^\infty{\int_0^{+\infty}{\mathrm d}\tau \frac{\mu}{\tau}e^{\left(
\frac{k^2}{2\mu}
-n\omega_1-m\omega_2\right)\tau-\frac{\mu\rho^2}{2\tau}}
\phi_n^{(1)}(Z)\phi_m^{(2)}(Z)\phi_n^{(1)}(Z')\phi_m^{(2)}(Z')
}
  \nonumber\\
 &=&  -2\mu X_0(k,\rho)\phi_0^{(1)}(Z)\phi_0^{(2)}(Z)\phi_0^{(1)}(Z')\phi_0^{(2)}(Z')\nonumber\\
  &&
  -\!\int_0^{+\infty}\!\mathrm{d}\tau\frac{\mu}{\tau}e^{
  \frac{k^2}{2\mu}
  -\frac{\mu\rho^2}{2\tau}}\left[
  g_{\mathrm{1D}}^{(1)}\left(Z,Z',\tau\right)
  g_{\mathrm{1D}}^{(2)}\left(Z,Z',\tau\right)
 - \phi_0^{(1)}(Z)\phi_0^{(2)}(Z)\phi_0^{(1)}(Z')\phi_0^{(2)}(Z')
  \right],
  \label{eq.new-omega2}
\end{eqnarray}
where $g^{(j)}_{\mathrm{1D}}\left(Z,Z',\tau\right)$ $(j=1,2)$$(j=1,2)$ denotes the imaginary-time Green’s function (propagator) of the 1D harmonic confinement for atom $j$:
\begin{eqnarray}
  g^{(j)}_{\mathrm{1D}}\left(Z,Z',\tau\right)
  &=&{\sum_{n=0}^{+\infty}\phi^{(j)}_n(Z)\phi^{(j)}_n(Z')\exp(-n\omega_j\tau)}\nonumber\\
  &=&\sqrt{\frac{m_j\omega_je^{\omega_j \tau}}{2\pi\sinh(\omega_j\tau)}}
  \exp\left\{
  -\frac{m_j\omega_j\left[
  (Z^2+Z^{\prime 2})\cosh(\omega_j\tau)
  -2ZZ^\prime
  \right]}{2\sinh(\omega_j\tau)}
  \right\}.
  \label{eq.imag-g1d}
\end{eqnarray}

Now we calculate $\Omega_{k=0}(0,Z,Z,Z')$.
Note that we cannot obtain $\Omega_{k=0}(0,Z,Z,Z')$
by directly taking $k=0$ and $\rho=0$ in  the expression (\ref{eq.new-omega2}) of  $\Omega_{k}(\rho,Z,Z,Z')$. That is because both the term $X_0(k,\rho)$ and the integration of this expression diverges in these two limits. 
To solve this problem, we 
re-write Eq.~(\ref{eq.new-omega2}) as
\begin{eqnarray}
 && \Omega_{k}(\rho,Z,Z,Z')\nonumber\\
  &=&
  \phi_0^{(1)}(Z)\phi_0^{(2)}(Z)\phi_0^{(1)}(Z')\phi_0^{(2)}(Z')
  2\mu\Big[K_0(\rho/\ell_z)-X_0(k,\rho)\Big] \nonumber\\
    &
  -&\int_0^{+\infty}\mathrm{d}\tau\frac{\mu}{\tau}e^{-\frac{\mu\rho^2}{2\tau}}\left[e^{\frac{k^2}{2\mu} \tau}g_{\mathrm{1D}}^{(1)}\left(Z,Z',\tau\right)
  g_{\mathrm{1D}}^{(2)}\left(Z,Z',\tau\right)-\left(e^{\frac{k^2}{2\mu} \tau}-e^{-\frac{\tau}{2\mu\ell_z^2}}\right) \phi_0^{(1)}(Z)\phi_0^{(2)}(Z)\phi_0^{(1)}(Z')\phi_0^{(2)}(Z')\right].\label{eq.new-omega3}
\end{eqnarray}
We can obtain
$\Omega_{k=0}(0,Z,Z,Z')$ by taking $k=0$ and $\rho=0$ for
 Eq.~(\ref{eq.new-omega3}). Explicitly, we obtain
\begin{eqnarray}
  \Omega_{k=0}(0,Z,Z,Z')&=&-2\mu\gamma \phi_0^{(1)}(Z)\phi_0^{(2)}(Z)\phi_0^{(1)}(Z')\phi_0^{(2)}(Z')
  \nonumber\\
  &&
   -\int_0^{+\infty}\mathrm{d}\tau\frac{\mu}{\tau}\left[g_{\mathrm{1D}}^{(1)}\left(Z,Z',\tau\right)
  g_{\mathrm{1D}}^{(2)}\left(Z,Z',\tau\right)
  -
  \left(1-e^{-\frac{\tau}{2\mu\ell_z^2}}\right) \phi_0^{(1)}(Z)\phi_0^{(2)}(Z)\phi_0^{(1)}(Z')\phi_0^{(2)}(Z')\right].  \nonumber\\
  \label{eq.new-omega4}
\end{eqnarray}
Notice that the integral in Eq.~(\ref{eq.new-omega4}) converges.
Similarly, $\Omega^{(2)}(0,Z,Z,Z')$ defined in (\ref{omega2}) is obtained as
\begin{eqnarray}
  \Omega^{(2)}(0,Z,Z,Z')= -\int_0^{+\infty}\mathrm{d}\tau\frac{1}{2}\left[
  g_{\mathrm{1D}}^{(1)}\left(Z,Z',\tau\right)
  g_{\mathrm{1D}}^{(2)}\left(Z,Z',\tau\right)
  -
  \phi_0^{(1)}(Z)\phi_0^{(2)}(Z)\phi_0^{(1)}(Z')\phi_0^{(2)}(Z')
  \right].\label{eo2}
\end{eqnarray}

Now we determine the value of $b_0$ by deriving the asymptotic behavior of $\Omega_{k=0}(\rho,Z,Z,Z')$ in the limit $\rho\to 0$ and $Z-Z^{\prime}\to 0$.
According to the expression~(\ref{eq.new-omega3}) of $\Omega_{k=0}(\rho,Z,Z,Z')$, when $\rho\rightarrow 0$ the integration of this expression is mainly contributed by the to-be-integrated function in the limit $\tau\to 0$.
On the other hand, in the $\tau\to 0$ limit we have
\begin{eqnarray}
  g_{\mathrm{1D}}^{(j)}\left(Z,Z',\tau\right)&\simeq&\sqrt{\frac{m_j}{2\pi\tau}}e^{-\frac{m_j(Z-Z^{\prime})^2}{2\tau}},\ \ \ 
  (j=1,2).
\end{eqnarray}
Using this result, we find that
the asymptotic behavior of $\Omega_{k=0}(\rho,Z,Z,Z')$ in the limit $\rho\to 0$ and $Z-Z^{\prime}\to 0$ is 
\begin{eqnarray}
  \Omega_{k=0}(\rho,Z,Z,Z')&\simeq&-\frac{\mu}{\pi}\frac{\sqrt{m_1m_2}}{\mu\rho^2+(m_1+m_2)(Z-Z^{\prime})^2}.
\end{eqnarray}
which gives
\begin{eqnarray}
b_0=\frac{m_1+m_2}{\sqrt{m_1m_2}}.
\end{eqnarray}

Finally, Eqs.~(\ref{eq.eta}-\ref{eq.LS-omega2}\label{sec.app-b})  reduce to
\begin{eqnarray}
  \frac{\mu}{2}\hat{S_0}\big[\eta^{(0)}(Z)\big]&=&\phi_0^{(1)}(Z)\phi_0^{(2)}(Z) + \frac{1}{2}\mathbb{Z}\int \Omega_{k=0}(0,Z,Z,Z') \eta^{(0)}\left(Z'\right)\mathrm{d}Z', \label{ne1} \\[5pt]
    \frac{\mu}{2}\hat{S_0}\big[\eta^{(2)}(Z)\big]+\frac{\mu}{2}R_{\rm3D} \eta^{(0)}(Z)&=&\frac{1}{2}\int \Omega^{(2)}(0,Z,Z,Z') \eta^{(0)}\left(Z'\right)\mathrm{d}Z' + \frac{1}{2}\mathbb{Z}\int \Omega_{k=0}(0,Z,Z,Z') \eta^{(2)}\left(Z'\right)\mathrm{d}Z',
  \label{ne2}
\end{eqnarray}
where the  Hadamard finite integral $\mathbb Z$ is defined in Eq.~(\ref{ehad2}) with $b_0=(m_1+m_2)/\sqrt{m_1m_2}$, and the keernels $\Omega_{k=0}(0,Z,Z,Z')$ and $\Omega^{(2)}(0,Z,Z,Z')$ are given by Eq.~(\ref{eq.new-omega4}) and Eq.~(\ref{eo2}), respectively. 

\section{Calculation for the Bound State Energy $E_b$}
\label{appbs}

\subsection{Summary of the Approach}

Our approach for calculating the bound-state energy
is also straightforwardly generalized from Ref.~\cite{2024}.
Specifically, we numerically solve the integral equation \eqref{nbound1} (whose meaning is clarified below) to obtain the eigenvalue $E_b$ and the corresponding function $\eta_b(Z)$ up to a normalization constant.  
In the following subsections, we derive Eq.~\eqref{nbound1} explicitly.
The following subsections present detailed derivations of Eqs.~(\ref{nbound1},  \ref{nboundEb}) and demonstrate the validity of the above steps.
The mathematical symbols in this appendix, except for those newly defined, are defined the same as in the main text and Appendices~\ref{a3d}, \ref{apps}.

\subsection{Derivation of Eq.\eqref{nbound1}}

Similar to the scattering state, the bound state of our system also 
lies in the $L_z=0$ subspace. As a result, the wave function $\Phi_b(\rho,z_1,z_2)$ is independent of the azimuthal angle of $\bm\rho$.  
This bound-state wave function satisfies the Schrödinger equation
\begin{eqnarray}\label{bound}
	{\hat H}\Phi_b\left(\rho, z_1, z_2\right)=E_b\Phi_b\left(\rho, z_1, z_2\right),
\end{eqnarray}
where $E_b<(\omega_{1}+\omega_{2})/2$.  
Equation~\eqref{bound} can therefore be recast into a LSE–type integral equation:
\begin{align}
	\Phi_b\left(\rho, z_1, z_2\right)= \int_{-\infty}^{+\infty}dz_1'\int_{-\infty}^{+\infty}dz_2'\int_0^{\infty}d\rho^\prime
	G_{E_b}\bigg(\rho, z_1, z_2 ; \rho^\prime, z_1^{\prime}, z_2^{\prime}\bigg) D_b\left(\rho^\prime, z_1^{\prime}, z_2^{\prime}\right),\label{eq.LS-bound}
\end{align}
where $G_E$ is defined in Eq.~\eqref{eq.G4d}, and
\begin{equation}
	D_b\left(\rho^\prime, z_1^\prime,z_2^\prime\right) ={\frac{2 \hat{A}_{\mathrm{3D}}\big(i\sqrt{2\mu |E_b|}\big)}{\mu}}
	\delta(z_1^\prime-z_2^\prime)\delta(\rho^\prime)
	\frac{1}{\rho^\prime}\frac{\partial}{\partial \rho^\prime}\bigg[\rho^\prime\Phi_b\big({\rho^\prime},z_1^\prime,z_2^\prime\big)\bigg],
\end{equation}
where the operator $\hat{A}_{\mathrm{3D}}(i\sqrt{2\mu |E_b|})$ is given in Eqs.~(\ref{a3d1}, \ref{a3d2}).
After integrating out the delta function in $D_b\left(\rho^\prime, z_1^\prime,z_2^\prime\right)$, we can reexpress Eq. \eqref{eq.LS-bound} as
\begin{eqnarray}
	\Phi_b(\rho, z_1, z_2)= \frac{1}{2}\int_{-\infty}^{+\infty}\mathrm{d}Z'\Omega_b\left(\rho,z_1, z_2,Z'\right)\eta_b(Z'),\label{lseb}
\end{eqnarray}
where the functions $\eta_b(Z)$ and $\Omega_b(\rho,z_1, z_2,Z') $ are given by
\begin{eqnarray}\label{defineetabound}
		\eta_b(Z)=\ \lim_{\rho\to 0}\frac{2
		\hat{A}_{\mathrm{3D}}(i\sqrt{2\mu |E_b|})
		}{\mu}\frac{\partial}{\partial \rho}\left[\rho\cdot\Phi_b(\rho,Z,Z)\right],
\end{eqnarray}
with
\begin{eqnarray}
	\Omega_b(\rho,z_1,z_2,Z')
	=-
	\sum_{m,n=0}^{\infty}\phi_m^{(1)}(z_1)\phi_n^{(2)}(z_2)\phi_m^{(1)}(Z^{\prime})\phi_n^{(2)}(Z^{\prime})\cdot2\mu K_{0}(\kappa_{m,n}\rho),\label{eq.omegabound}
\end{eqnarray}
and $\kappa_{m,n}=\sqrt{-2\mu (E_b -E_m-E_n)}$.
By combining Eq.~(\ref{defineetabound}) and the Bethe-Peierls boundary condition satisfied by the wave function $\Phi_b(\rho,z_1,z_2)$,  we obtain
\begin{eqnarray}
	\Phi_b(\rho,Z,Z)=-\frac{\mu}{2}\left(\frac{1}{\rho}\eta_b(Z) -\frac{1}{\hat{A}_{\mathrm{3D}}(i\sqrt{2\mu |E_b|})}\eta_b(Z) \right) + \mathcal{O}(\rho).\label{eqetabound}
\end{eqnarray}
This result, together with Eq.~(\ref{a3d2}), further leads to
\begin{eqnarray}
\lim_{\rho\rightarrow 0}
\bigg[
\Phi_b(\rho,Z,Z)+\frac{\mu}{2}\frac{1}{\rho}\eta_b(Z)
\bigg]
=\frac{\mu}{2}\left[\frac{1}{a_{\rm{3D}}}+R_{\rm{3D}}2\mu E_b-R_{\rm{3D}}2\mu\hat{H}_{\rm com}\right]\eta_b(Z).\label{eq.eqhatbound}
	\end{eqnarray}
Substituting Eq. \eqref{lseb} into the above result, we obtain the integral equation
for $\eta_b(Z)$ as
\begin{eqnarray}
	\frac{\mu}{2}\left[\frac{1}{a_{\rm{3D}}}+R_{\rm{3D}}2\mu E_b-R_{\rm{3D}}2\mu\hat{H}_{\rm com}\right]\eta_b(Z)=\frac{1}{2}\lim\limits_{\rho\to0}\mathcal{I}_{b}(\rho,Z),\label{eq.eqhatb} 
\end{eqnarray}
where the functions $\mathcal{I}_{b}(\rho,Z)$ are defined as:
\begin{eqnarray}
	\mathcal{I}_{b}(\rho,Z)&\equiv&\int_{-\infty}^{+\infty}\Omega_{b}(\rho,Z,Z,Z')\eta_{b}(Z')\mathrm{d}Z' + \frac{\mu}{\rho}\eta_{b}(Z),\ \ \ \ (j=0,2).\label{eq.Ib}
\end{eqnarray}
Similar as in the above subsection, we can further re-express the terms $\lim_{\rho\to0} \mathcal{I}_{b}(\rho,Z)$ of Eqs. (\ref{eq.eqhatb}) as Hadamard finite part integral:
\begin{eqnarray}
	\lim_{\rho\to 0}\mathcal{I}_{b}(\rho,Z)  &=& \mathbb{Z}\int \Omega_{b}(\rho,Z,Z,Z')\eta_{b}(Z')\mathrm{d}Z',\nonumber\\
	&\equiv&\lim_{\epsilon\to 0}\left[\int_{-\infty}^{Z'-\epsilon} \Omega_{b}(0,Z,Z,Z')\eta_{b}(Z')\mathrm{d}Z' + \int_{Z'+\epsilon}^{+\infty}\Omega_{b}(0,Z,Z,Z')\eta_{b}(Z')\mathrm{d}Z' + \frac{2}{b_0}\frac{\mu}{\pi}\frac{\eta_b(Z)}{\epsilon} \right].
	\label{ehadbound}
\end{eqnarray}
Moreover, we  rewrite $\Omega_b(0,Z,Z,Z')$ using the Laplace representation of the modified Bessel function (i.e., Eq.~\eqref{eq.laplace-k}):
 \begin{eqnarray}
 \Omega_{b}(\rho,Z,Z,Z')=-\sum_{m,n=0}^{\infty}\int_0^{+\infty}\mathrm{d}\tau\frac{\mu}{\tau}e^{-\frac{\mu\rho^2}{2\tau}}\left[e^{(E_b-E_m-E_n)\tau}\phi_m^{(1)}(Z)\phi_n^{(2)}(Z)\phi_m^{(1)}(Z^{\prime})\phi_n^{(2)}(Z^{\prime}) \right].\label{eq.new-omegabound}
 \end{eqnarray}
Using the above results and the method similar to the one in Appendix \ref{sec.app-b}, we can 
obtain $b_0=(m_1+m_2)/\sqrt{m_1m_2}$, and
derive the homogeneous integral equation for $E_b$
and $\eta_b(Z)$:
\begin{eqnarray}
\frac{\mu}{2}\left[\frac{1}{a_{\rm{3D}}}+R_{\rm{3D}}2\mu E_b-R_{\rm{3D}}2\mu\hat{H}_{\rm com}\right]\eta_b(Z)=\frac{1}{2}\mathbb{Z}\int F_b(E_b,Z,Z') \eta_b\left(Z'\right)\mathrm{d}Z', \label{nbound1} 
\end{eqnarray}
where  the kernel  $F_b(E_b,Z,Z')$ is given by
\begin{eqnarray}
F_b(E_b,Z,Z')=-\int_{0}^{+\infty}\mathrm{d}\tau\frac{\mu}{\tau}e^{(E_b-\frac{\omega_{1}+\omega_{2}}{2})\tau} g_{\mathrm{1D}}^{(1)}\left(Z,Z',\tau\right) g_{\mathrm{1D}}^{(2)}\left(Z,Z',\tau\right)\label{nboundEb} 
\end{eqnarray}
with  $g^{(j)}_{\mathrm{1D}}\left(Z,Z',\tau\right)$  being defined in Eq.~\eqref{eq.imag-g1d}.

\end{widetext}

\nocite{*}


\begin{thebibliography}{99}
\bibitem{2D}
N. D. Mermin and H. Wagner, Phys. Rev. Lett. {\bf 17}, 1133 (1966).

\bibitem{2DBKT1}
V. L. Berezinsky, Sov. Phys. JETP 32, 493 (1971).

\bibitem{2DBKT2}
 J. M. Kosterlitz and D. J. Thouless, J. Phys. C 6, 1181 (1973).

\bibitem{BoseBKT}
Z. Hadzibabic, P. Kr\"uger, M. Cheneau, B. Battelier, and J. Dalibard,
Nature {\bf 441}, 1118 (2006).

\bibitem{Q2D1}
I. Bloch, J. Dalibard, and W. Zwerger, Rev. Mod. Phys. {\bf 80}, 885 (2008).

\bibitem{Q2D2}
Z. Hadzibabic and J. Dalibard, Rivista Del Nuovo Cimento, \href{http://dx.doi.org/10.1393/ncr/i2011-10066-3}{{\bf 34}, 389 (2011).}

\bibitem{}
J. M. Kosterlitz and D. J. Thouless, J. Phys. C {\bf 6}, 1181 (1973).

\bibitem{2DBose1}
N. Prokof’ev, O. Ruebenacker, and B. Svistunov, Phys. Rev. Lett. {\bf 87}, 270402 (2001).

\bibitem{2DBose2}
P. Kr\"uger, Z. Hadzibabic, and J. Dalibard, Phys. Rev. Lett. {\bf 99}, 040402 (2007).

\bibitem{2DBose3}
R. Desbuquois et al., Nat. Phys. {\bf 8}, 645 (2012).

\bibitem{2DBose4}
R. J. Fletcher et al., Phys. Rev. Lett. {\bf 114}, 255302 (2015).

\bibitem{2DFeimi1}
M. Feld et al., Nature {\bf 480}, 75 (2011).

\bibitem{2DFeimi2}
M. G. Ries et al., Phys. Rev. Lett. {\bf 114}, 230401 (2015).

\bibitem{2DFeimi3}
K. Fenech et al., Phys. Rev. Lett. {\bf 116}, 045302 (2016).

\bibitem{2DFeimi4}
C. Daix et al., Phys. Rev. Lett. {\bf 136}, 153402 (2026).

\bibitem{Petrov}
D. S. Petrov and G. V. Shlyapnikov, Phys. Rev. A {\bf 64}, 012706 (2001).

\bibitem{Duan2006}
J. P. Kestner and L.-M. Duan, Phys. Rev. A {\bf 74}, 053606 (2006).

\bibitem{Hu}
H. Hu, B. C. Mulkerin, U. Toniolo, L. He, and X.-J. Liu, Phys. Rev. Lett. {\bf 122}, 070401 (2019).

\bibitem{heteronuclear1}
C. H. Wu et al.,  Phys. Rev. Lett. {\bf 109}, 085301 (2012). 

\bibitem{heteronuclear2}
G. Barontini et al.,  Phys. Rev. Lett. {\bf 103}, 043201 (2009).

\bibitem{heteronuclear3}
R. Pires et al., Phys. Rev. Lett. {\bf 112}, 250404 (2014). 

\bibitem{heteronuclear4}
J. Levinsen and M. M. Parish, Annu. Rev. Cold At. Mol. \href{https://doi.org/10.1142/9561}{{\bf 3}, 71 (2015).}

\bibitem{heterFemi1}
K. B. Gubbels and H. T. C. Stoof, Phys. Rep. {\bf 525}, 255 (2013)

\bibitem{heterFemi2}
J. Wang et al., Sci. Rep. {\bf 7}, 39783 (2017).


\bibitem{heteronuclear2D}
J. Ulmanis et al., Nat. Sci. Rev. {\bf 3}, 174 (2016).

\bibitem{solve2D1}
D. S. Petrov, M. Holzmann and G. V. Shlyapnikov, Phys. Rev. Lett.  {\bf 84}, 2551 (2000).

\bibitem{solve2D2}
J. Levinsen and M. M. Parish, Phys. Rev. Lett. {\bf 110}, 055304 (2013).

\bibitem{2024} F. Yang, R. Du, R. Qi, and P. Zhang,  Phys. Rev. A {\bf 110}, 033318 (2024).

\bibitem{LHY1}
K. Huang and C. N. Yang, Phys. Rev. 105, 767 (1957).

\bibitem{LHY} D. Xiao, R. Zhang, and P. Zhang, Phys. Rev. Research {\bf 4}, 013112 (2022).


\bibitem{space} The
scattering state \(\Psi_{\bm k}^{(+)}({\bm \rho},z_1,z_2)\) discussed
in the preceding section belongs precisely to the \({\cal P}=+1\) subspace.

\end{thebibliography}

\end{document}